\documentclass[aps,prd,reprint,amsmath,amssymb,superscriptaddress,floatfix]{revtex4}
\usepackage{graphicx}
\usepackage{verbatim} 
\usepackage{amsfonts}
\usepackage{amssymb}
\usepackage{rotating}
\usepackage{booktabs}
\usepackage{xcolor}
\usepackage{soul}
\usepackage{color}
\usepackage{slashed}
\usepackage{multirow}
\usepackage{makecell}
\usepackage{epsf}
\usepackage{ulem}
\usepackage{cancel}
\usepackage{color,bm}
\usepackage[colorinlistoftodos]{todonotes}
\usepackage{diagbox}
\usepackage[colorlinks=true,citecolor=cyan,urlcolor=blue,bookmarks=true,bookmarks=true,bookmarksopen=true,bookmarksnumbered=true,bookmarksopenlevel=3]{hyperref}
\usepackage{subfigure}
\definecolor{airforceblue}{rgb}{0.36, 0.54, 0.66}
\definecolor{steelblue}{rgb}{0.27, 0.51, 0.71}
\definecolor{amber}{rgb}{1.0, 0.49, 0.0}

\begin{document}

\title{$Z_5$ two-component dark matter in the Type-II seesaw mechanism}

\author{\textsc{XinXin Qi}}
\email{qxx@dlut.edu.cn}
\affiliation{Institute of Theoretical Physics, School of Physics, Dalian University of Technology, No.2 Linggong Road, Dalian, Liaoning, 116024, P.R.China }
\author{\textsc{Hao Sun}}
\email{haosun@dlut.edu.cn}
\affiliation{Institute of Theoretical Physics, School of Physics, Dalian University of Technology, No.2 Linggong Road, Dalian, Liaoning, 116024, P.R.China }

\begin{abstract}
We consider the  $Z_5$ two-component dark matter model within the framework of the Type-II seesaw mechanism. Due to the new annihilation processes related to triplets, the light component cannot necessarily be dominant in the dark matter relic density, which is different from the $Z_5$ two-component dark matter model in the SM. 
  The model is considered to explain the excess of electron-positron flux
measured by the AMS-02 Collaborations in this work, which is encouraged by the decay of
the triplets arising from dark matter annihilations in the Galactic halo. We discuss the cases of the light and heavy components determining dark matter density within a viable parameter space satisfying relic density and direct detection constraints, and 
by fitting the antiproton spectrum observed in the PAMELA and AMS experiments, we find that the parameter space is flexible and the electron-positron flux excess can be obtained in both cases with the mass of two dark matter particles being larger than that of the triplets’.
\vspace{0.5cm}

\vspace{0.5cm}
\end{abstract}
\maketitle
\setcounter{footnote}{0}
\section{Introduction}
\label{sec:intro}
The Standard Model (SM) has achieved great success with its high accuracy and prediction for particle physics. However, dark matter and neutrino mass are still two open problems that SM can not explain. There is no space for dark matter in the SM while a great deal of astronomical evidence has indicated the existence of dark matter \cite{Zwicky:1933gu,Allen:2011zs,Frenk:2012ph,WMAP:2012nax}. On the other hand, neutrinos are massless with the absence of right-handed neutrino in the SM  which is also contradicted with the neutrino oscillation experiments \cite{ParticleDataGroup:2014cgo}.

From the view of particle physics, one of the most attractive schemes to solve dark matter problem is WIMPs (weak interaction massive particles), where the stable dark matter particle is at a weak scale and the observed dark matter relic density is generated via the so-called Freeze-out mechanism\cite{Chiu:1966kg}. In addition, scenarios in which two or more different particles contribute to the dark matter density— multi-component dark matter model are also possible \cite{Belanger:2021lwd,Alguero:2023zol,Belanger:2022esk,Bhattacharya:2016ysw,Bhattacharya:2019fgs,Bhattacharya:2022qck,Bhattacharya:2022wtr}, where the total relic density of two or more dark matter particles contributes to the experiment results.  For the different constituents of dark matter in the multi-component dark matter model, the corresponding density can be generated not just via the Freeze-out mechanism but also the Freeze-in mechanism \cite{Hall:2009bx}. Correspondingly, the components of dark matter can be WIMPs, FIMPs( Feebly interaction massive particles), or both. Among the multi-component dark matter models, the scalar fields that are
simultaneously stabilized by a single $Z_N$ symmetry as dark matter particles are particularly appealing\cite{Deshpande:1977rw,Ma:2006km,Barbieri:2006dq,LopezHonorez:2006gr,Belanger:2014bga,Belanger:2012vp,DiazSaez:2022nhp}, these complex scalars are SM singlets but have different charges under the discrete symmetry. This symmetry, in turn, could be a remnant of a spontaneously broken $U(1)$ gauge symmetry and thus be related to gauge extensions of the SM \cite{Martin:1992mq,Krauss:1988zc}. 

As for the neutrino mass problem, one can introduce new particles to the SM to generate the tiny neutrino mass with the so-called seesaw mechanism\cite{Schechter:1980gr,Schechter:1981cv,Schechter:1981bd}. Generally speaking, we have three kinds of seesaw mechanisms: Type-I seesaw \cite{Minkowski:1977sc,Mohapatra:1979ia} (introducing new right-handed neutrinos with Majorana mass terms), Type-II seesaw \cite{Magg:1980ut,Lazarides:1980nt,Ashery:1981tq,Ma:1998dx,Konetschny:1977bn,Schechter:1980gr,Cheng:1980qt,Bilenky:1980cx}(introducing new triplet scalars) and Type-III seesaw \cite{Foot:1988aq,Ma:1998dn,Ma:2002pf}(introducing new fermion triplets). For the Type-II seesaw mechanism, we have a new triplet scalar $\Delta \equiv (\delta^{++},\delta^+,\delta^0)$ with a non-zero vacuum expectation value ($v_{\Delta}$). Since the triplet couples to SM gauge bosons and leptons directly, rich LHC phenomenology for a low seesaw scale can be found in the Type-II seesaw \cite{Akeroyd:2007zv,FileviezPerez:2008jbu,delAguila:2008cj,Melfo:2011nx}. What's more, the light triplet 
can modify the  Higgs-to-Z+photon \cite{Djouadi:1996yq,Carena:2012xa,Chen:2013vi} and  Higgs-to-diphoton \cite{Melfo:2011nx,Arhrib:2011vc,Akeroyd:2012ms} decay rates, which are correlated in the type-II seesaw model for most of the viable parameter space \cite{BhupalDev:2013xol,Chen:2013dh}.

In this work, we consider the $Z_5$ two-component dark matter in the Type-II seesaw mechanism. The case of $Z_5$ two-component dark matter in the SM has been discussed in \cite{Belanger:2020hyh,Qi:2023egb}, where two singlet scalars carrying $Z_5$ charge are introduced to the SM with the Higgs-portal couplings as dark matter, and we have new trilinear and quartic couplings of the two scalars due to the $Z_5$ symmetry, which can induce annihilation, semi-annihilation and conversion processes related to dark matter density. One feature of such a model is that the entire range of dark matter masses is allowed, and the dark matter density is always dominated by the lighter component. In this work, we have new couplings of dark matter with the triplet scalar so that new annihilation processes of the two singlet scalars are possible, which means the lighter component does not necessarily determine dark matter density. On the other hand, when dark matter is embedded in the Type-II seesaw,  
its interactions with the triplet scalar field $\Delta$ which dominantly decays to leptonic final states for a small triplet vacuum expectation value $v_{\Delta}<0.1$ MeV can contribute to fit the excess of positron-electron flux according to the AMS-02 data for the leptophilic nature of dark matter. Concretely speaking, the introduced triplet in the type-II seesaw mechanism can play an important role in exploring the observed excess of cosmic rays observed in the electron-positron flux measured by the AMS-02 \cite{AMS:2014xys}, Fermi-LAT \cite{Fermi-LAT:2017bpc}, and DAMPE \cite{DAMPE:2017fbg}experiments, arising
from the leptonic decays of such a triplet during DM annihilation \cite{Dev:2013hka,Li:2017tmd,Li:2018abw}.  The W and Z boson pair from s-channel DM annihilation with subsequent decay may lead to an inappropriate antiproton spectrum measured by AMS and PAMELA in cosmic rays, which constrains the coupling between SM Higgs and dark matter. Note that this model requires a larger  boost factor (BF)  of order $10^3 - 10^4$ in the DM annihilation rate to explain the observed positron excess \cite{Dev:2013hka}, which may come from small-scale inhomogeneities in the dark matter distribution which cannot be excluded even with the highest resolution numerical simulations available at present\cite{Kuhlen:2012ft},or due to the so-called Breit-Wigner enhancement mechanism in particle physics \cite{Ibe:2008ye,March-Russell:2008klu,Guo:2009aj}. Related discussion about cosmic ray excess in the Type-II seesaw mechanism can be found in \cite{Dev:2013hka,Qi:2021rpa}. In this paper, we consider the two-component dark matter case in the Type-II seesaw mechanism where both components of the dark matter can contribute to the possible positron-electron excess. As we mentioned above, either the light or the heavy component can be dominant in the model, and we will have different results to fit the observed positron-electron excess in both cases.

The paper is arranged as follows, in Sec.~\ref{sec:2}, we give 
the model framework of $Z_5$ two-component dark matter in the Type-II seesaw mechanism. In Sec.~\ref{sec:3}, we discuss the dark matter phenomenology in the model. In Sec.~\ref{sec:4}, we consider the anti-proton spectrum and electron/positron excess in our framework, and we give a summary in the last part.
\section{Model description}
\label{sec:2}
 The Lagrangian of two-component dark matter in the Type-II seesaw mechanism can be given as follows:
 \begin{eqnarray}
 \mathcal{L}= \mathcal{L_{SM}} + \mathcal{L}_{type-II} + \mathcal{L}_{DM}
 \end{eqnarray}
 where $\mathcal{L}_{SM}$ is the Lagrangian of SM, $\mathcal{L}_{type-II}$ are terms related to Type-II seesaw and $\mathcal{L}_{DM}$ is the Lagrangian related to dark matter. Concretely speaking,

\begin{eqnarray} \nonumber
\mathcal{L}_{type-II} &=& M^2_{\Delta} \mathrm{Tr}(\Delta^\dagger\Delta) +[\mu_1 (H^Ti{\sigma}^2\Delta^{\dagger}H)+ \mathrm{h.c.}]
  + \lambda_1(H^\dagger H) \mathrm{Tr}(\Delta^\dagger \Delta) + \lambda_2(\mathrm{Tr}\Delta^\dagger\Delta)^2 +
 \lambda_3 \mathrm{Tr}(\Delta^\dagger \Delta)^2+\lambda_4 H^\dagger\Delta\Delta^\dagger H \\
\end{eqnarray}
$H$ and $\Delta$ are labels of the Higgs doublet and the triplet scalar respectively which are represented as
\begin{eqnarray}
H&=&\begin{pmatrix}{G^+}\\{\frac{1}{\sqrt{2}}(v_0+h+iG^0)}
\end{pmatrix}\ , \\
\Delta&=&\begin{pmatrix}{\frac{1}{\sqrt{2}}\delta}^{+} & {\delta}^{++}\\\frac{1}{\sqrt{2}}(v_{\Delta}+{\delta}^0+i\eta^{0}) &-\frac{1}{\sqrt{2}}\delta^{+}
\end{pmatrix} \ \ {\mathrm{or}}\ \
\begin{pmatrix}
\delta^{++}\\
\delta^{+}\\
 \frac{1}{\sqrt{2}}(v_\Delta+\delta^0+i \eta^0)
\end{pmatrix}
\end{eqnarray}
with $v_0$ ($v_{\Delta}$) is the VEV of $H$ ($\Delta$).
$G^0$, and $G^\pm$ are the Goldstone bosons that are eaten up to give mass to SM gauge bosons.

For the dark matter part, we have:
 \begin{align}
\label{eq:Z5lag}
 \mathcal{L}_{DM}&=\mu_{1}^2|\phi_1|^2+\lambda_{41}|\phi_1|^4+\lambda_{s1h}|H|^2|\phi_1|^2 \,+\mu_{2}^2|\phi_2|^2+\lambda_{42}|\phi_2|{^4}+\lambda_{s2h}|H|^2|\phi_2|^2+ \lambda_{s1d}|\phi_1|^2 \mathrm{Tr}(\Delta^{\dagger}\Delta) \nonumber \\
  &+ \lambda_{s2d}|\phi_2|^2 \mathrm{Tr}(\Delta^{\dagger}\Delta) +\lambda_{412}|\phi_1|^2|\phi_2|^2 +\frac{1}{2}\left[\mu_{s1}\phi^2_1\phi_2^{*} + \mu_{s2}\phi_2^2\phi_1 +\lambda_{31}\phi _1^3 \phi _2+\lambda_{32}\phi _1 \phi _2^{*3} + \text{h.c.}\right] 
\end{align}
 where $\phi_1$ and $\phi_2$ are dark matter carrying $Z_5$ charge with $X_{\phi_1}=1/5$ and $X_{\phi_2}=2/5$, and we have the  transform of $SM \to SM$, $\phi_1 \to e^{i2\pi/5}\phi_1$ and $\phi_2 \to e^{i4\pi/5} \phi_2$ under the $Z_5$ symmetry. The new trilinear and quartic couplings denoted $\mu_{si}$ and $\lambda_{3i}~({i= 1,2})$ respectively.
 
By solving the minimal condition of $\partial{\cal{V}}(v_0,v_{\Delta},0)/\partial{v_{\Delta}}=0$ 
and $\partial{\cal{V}}(v_0,v_{\Delta},0)/\partial{v_0}=0$ under the condition $v_{\Delta} \ll v_0$, we can get
\begin{eqnarray}\label{Gv2}
 v_{0}=\sqrt{\frac{-\mu^2}{\lambda}},\ \ v_{\Delta}\approx \frac{\mu_1 v^2_0}{\sqrt{2}(M^2_{\Delta}+\frac{\lambda_1+\lambda_4}{2}v^2_0)}.
\end{eqnarray}

The value of $\mu_1$ is small in the scheme of $\mu_1 \sim v_{\Delta}$ 
so that we can neglect the associated contribution to DM annihilation.
For the doubly charged scalar masses, we have 
\begin{eqnarray}
M^2_{\delta^{\pm\pm}} = -v^2_\Delta \lambda_3 - \frac{\lambda_4}{2} v^2_0 + \frac{\mu_1}{\sqrt{2}} \frac{v^2_0}{v^2_\Delta} .
\end{eqnarray}
Here and in the following, without confusion, we use the flavor eigenstate symbol to label its mass eigenstate.
The mass squared matrix for the singly charged field can be diagonalized, with one eigenvalue zero corresponding to the charged Goldstone
boson $G^\pm$ while the other corresponds to the singly charged Higgs boson $\delta^\pm$ which can be given by
\begin{eqnarray}
M^2_{\delta^\pm} = - \frac{v^2_0+2 v^2_\Delta}{4 v_\Delta} ( v_\Delta \lambda_4 - 2 \sqrt{2} \mu_1 ) .
\end{eqnarray}
When the neutral scalar mass matrice is diagonalized, one obtains two massive even-parity physical
states $h$ and $\delta^0$ with the masses:
\begin{eqnarray}
M^2_{h} &=& \frac{1}{2}(A + B - \sqrt{(A-B)^2+4 C^2}), \\
M^2_{\delta^0} &=& \frac{1}{2}(A + B + \sqrt{(A-B)^2+4 C^2}).
\end{eqnarray}
with 
\begin{eqnarray}
A &=& 2 v^2_0 \lambda, \\
B &=& 2 v^2_\Delta ( \lambda_2 + \lambda_3) + \frac{\mu_1}{\sqrt{2}}\frac{v^2_0}{v_\Delta}, \\
C &=& v_0 ( v_\Delta (\lambda_1+\lambda_4) - \sqrt{2} \mu_1 ) .
\end{eqnarray}
The pseudoscalar mass matrices lead to one massless Goldstone boson $G^0$ and one massive physical state $\eta^0$
\begin{eqnarray}
M^2_{\eta^0} = \frac{v^2_0 + 4 v^2_\Delta}{\sqrt{2}v_\Delta} \mu_1 .
\end{eqnarray}

From the relation listed above, we can write the coupling parameters as the function of the masses
\begin{eqnarray}
\mu_1 &=& \frac{\sqrt{2}v_\Delta}{v^2_0+4 v^2_\Delta} M^2_{\eta^0}, \\
\lambda &=& \frac{1}{2v^2_0} ( M^2_{h} \cos^2{\beta} + M^2_{\delta^0} \sin^2{\beta} ), \\
\lambda_4 &=& \frac{4}{v^2_0+4v^2_\Delta} M^2_{\eta^0} - \frac{4}{v^2_0 + 2 v^2_\Delta} M^2_{\delta^{\pm}}, \\
\lambda_3 &=& \frac{1}{v^2_\Delta} \left( \frac{-v^2_0}{v^2_0+4 v^2_\Delta}M^2_{\eta^0} + \frac{2v^2_0}{v^2_0+2 v^2_\Delta} M^2_{\delta^\pm} - M^2_{\delta^{\pm\pm}}  \right), \\
\lambda_2 &=& \frac{1}{v^2_\Delta} \left( \frac{\sin^2\beta M^2_{h} 
+ \cos^2\beta M^2_{\delta^0}}{2} + \frac{1}{2} \frac{v^2_0}{v^2_0+4 v^2_\Delta} M^2_{\eta^0} - \frac{2 v^2_0}{v^2_0+2 v^2_\Delta} M^2_{\delta^\pm} + M^2_{\delta^{\pm\pm}}     \right),\\
\lambda_1 &=& - \frac{2}{v^2_0 + 4 v^2_\Delta} M^2_{\eta^0} + \frac{4}{v^2_0 + 2 v^2_\Delta} M^2_{\delta^{\pm}} + \frac{\sin{2\beta}}{2v_0v_\Delta}( M^2_{h} - M^2_{\delta^0} ).
\end{eqnarray}
with the mixing angle $\beta$ satisfying
\begin{eqnarray}
\sin(2\beta) &=& \frac{4 v_0 \left[ -5 \left( 4 M^2_\Delta + 2 (M^2_{h}+M^2_{\delta^0}) + M^2_h \right) v^2_\Delta \left(v^2_0 + 4 v^2_\Delta\right) + M^2_{\eta} \left(4v^4_0 + 6 v^2_0 v^2_\Delta + 5 v^4_\Delta\right) \right]}
{5 (M^2_{h}-M^2_{\delta^0}) v_\Delta (4 v^2_0 + v^2_\Delta) (v^2_0 + 4 v^2_\Delta)}.~~~~~~~~ 
\end{eqnarray}
As for the Yukawa and neutrino mass terms, we follow the results of \cite{Qi:2021rpa}.
For the dark matter mass terms, we have: 
\begin{eqnarray}\label{Gv3}
m^2_1=\mu^2_1 + \frac{\lambda_{s1h}}{2} v^2_0 + \frac{\lambda_{s1d}}{2} v^2_\Delta.  \\
m^2_2= \mu^2_2 + \frac{\lambda_{s2h}}{2} v^2_0 + \frac{\lambda_{s2d}}{2} v^2_\Delta.
\end{eqnarray}
To guarantee the stability of dark matter $\phi_1$ and $\phi_2$, we can consider $m_1 < m_2 <2m_1$. Without loss of generality, we can assume that  $\phi_1 $ is heavier than $\phi_2 $ so that $m_2 < m_1 < 2m_2$ due to the symmetry of the Lagrangian. We consider the case $m_1 < m_2 <2m_1$ in the following discussion for simplicity.

According to \cite{FileviezPerez:2008jbu,Li:2018abw,FileviezPerez:2008wbg}, for $v_{\Delta} \leq 10^{-4}$ GeV, the decays of the doubly charged Higgs boson are dominantly a same-sign dilepton.
For numerical purposes, we have chosen $v_{\Delta} =1 $ eV satisfying the experiment constraints. However, our results are independent of the exact value of $v_{\Delta}$ as long as $v_{\Delta} \leq 0.1$ MeV so that the leptonic branching ratio for the $\Delta$’s is
almost $100\%$.
\section{Theoritical constraints}
In this part, we consider the theoritical constraints on the parameter space including perturbativity, perturbative unitarity and vacuum stability.
\subsection{Perturbativity}
To illustrate the theoretical bounds from the perturbativity behavior of the dimensionless scalar quartic couplings, we follow the definitions in Refs.~\cite{Belanger:2014bga,Lerner:2009xg}. As to the
case of an unrotated basis, the vertices from the potential
must be less than $4\pi$ to make sure that the tree-level
contributions are larger than the one-loop level quantum
corrections. This condition will give the constraints on the quartic couplings  in the potential, which are
\begin{eqnarray*}
|\lambda_1 + \lambda_4 |<4\pi,|\lambda_1+ \frac{\lambda_4}{2}|<4\pi, |6(\lambda_2+\lambda_3)|<4\pi,|2\lambda_2|<4\pi,|\sqrt{2}\lambda_3|<4\pi,|\lambda_{sid}|<4\pi,
\end{eqnarray*}
\begin{eqnarray}
|\lambda_{sih}|<4\pi|,|4\lambda_{4i}|<4\pi,|3\lambda_{3i}|<4\pi,|\lambda_{412}|<4\pi
\end{eqnarray}
where $i=1,2$.
\subsection{Perturbative unitarity}
The tree-level unitarity from two-body scalar-scalar
scattering processes gives another bound on the quartic couplings in the potential. When the collision energy $\sqrt{s}$
becomes larger, the processes will be dominated by the terms of
quartic contact interaction. Although the trilinear couplings
that are contributed to scattering should be included at finite
collision energy \cite{Hektor:2019ote,Schuessler:2007av}, for simplicity, we only calculate
the unitarity constraints with the following scenario:
$s \to +\infty$. The s-wave scattering amplitudes lie in the
perturbative unitarity limit, giving the constraint of the
scalar-scalar scattering S-matrix values: $|\mathrm{Re}\mathcal{M}_i| \leqslant \frac{1}{2}$.
The perturbative unitarity in the type-II seesaw model
has been studied by decomposing the matrix S by the
mutually unmixed sets of channels with definite charge and
CP states \cite{Arhrib:2011uy}. We extend the way of decomposing by
considering the $Z_5$ symmetry with
$X_{\phi_1}= 1/5$ singlet $\phi_1$ and $X_{\phi_2}=2/5$ singlet $\phi_2$ introduced in our model. The matrix S can be decomposed into seven submatrix blocks structured in terms of electric
charges and $Z_5$ charges in the initial/final states.

 In Appendix.~\ref{appA}, we display the initial/final states $E_i$ and the corresponding scattering submatrix $\mathcal{M}_i$. The corresponding eigenvalues $e^j_i$ of each submatrix are then calculated. The limit from perturbative unitarity on the potential’s  quartic couplings, i.e., $|\mathrm{Re}\mathcal{M}_i| \leqslant \frac{1}{2}$, infers $|e^j_i|\leqslant 8\pi$.
\subsection{Vacuum stability}
The vacuum stability requires the potential under the restriction,i.e,$\mathcal{V} > 0$ for large field
values. The quadratic and cubic terms in the scalar
potential can be ignored compared with the quartic term in this limit. In this section, we focus on the vacuum stability induced by dark matter scalars and ignore the
mixing quartic couplings with the SM Higgs as well as the triplet scalar for simplicity, and this scenario can work when the mixing quartic
couplings with the SM Higgs and triplet scalar take positive or small values as compared to couplings in the
dark matter fields. Therefore, the potential with vacuum stability can be given by
\begin{align}
\mathcal{V}_{\mathrm{DM}} = \frac{1}{4}\lambda_{41}\Phi_1^4 + \frac{1}{4}\lambda_{42}\Phi_2^4 +\frac{1}{4} \lambda_{412} \Phi_1^2 \Phi_2^2  + \frac{1}{4}\lambda_{31}\Phi_1^3\Phi_2 cos(3\theta_1 +\theta_2) +\frac{1}{4}\lambda_{32}\Phi_1\Phi_2^3 cos(\theta_1 - 3\theta_2).
\end{align}
 where we have parameterized the dark matter fields with $\phi_1 =\Phi_1 e^{i\theta_1}$ and $\phi_2 =\Phi_2e^{i\theta_2}$.
 
After minimizing the potential for $3\theta_1 +\theta_2$ and $\theta_1 - 3\theta_2$ with $\Phi_1 \neq 0$ and $\Phi_2 \neq 0$, the above potential has the form\cite{Qi:2023egb}, 
\begin{align}
\mathcal{V}_{\mathrm{DM}} = \frac{1}{4}\lambda_{41}\Phi_1^4 + \frac{1}{4}\lambda_{42}\Phi_2^4 +\frac{1}{4} \lambda_{412} \Phi_1^2 \Phi_2^2  - \frac{1}{4}\lambda_{31}\Phi_1^3\Phi_2 -\frac{1}{4}\lambda_{32}\Phi_1\Phi_2^3.
\end{align}
Taking $X\equiv \frac{\Phi_1}{\Phi_2}$, the vacuum stability conditions become
\begin{align}
\lambda_{41} >0 ,    \lambda_{42} >0
\end{align}
and 
\begin{align}\label{leq0}
f(X_{min}) > 0
\end{align}
with
\begin{align}
f(X_{min}) =  \frac{1}{4}\lambda_{41}X^4 -\frac{1}{4}\lambda_{31}X^3 
+\frac{1}{4} \lambda_{412} X^2 - \frac{1}{4}\lambda_{32}X +\frac{1}{4}\lambda_{42},  
\end{align}
where $X_{min}$ labels the global minimum when $f^{'}(X_{min})=0$. Then, solving $f^{'}(X_{min})=0$ by the condition $f^{''}(X_{min}) > 0$, we obtain the third condition (\ref{leq0}) as

\begin{align}
\lambda_{41}X_{min}^3 - \frac{3}{4}\lambda_{31}X^2 +\frac{1}{2}\lambda_{412}X-\frac{1}{4}\lambda_{32} = 0,\label{CUB}
\\
\lambda_{42} > \frac{1}{4}\lambda_{31}X^3_{min}+\frac{1}{2}\lambda_{412}X^2_{min}+\frac{3}{4}\lambda_{32}X_{min}.
\end{align}
where
\begin{align}
X_{ min} = \Bigg\{
\begin{array}{ll}
\left(
P+\sqrt{P^2 + Q^3}    
\right)^{1/3}
+\left(
P-\sqrt{P^2 + Q^3}    
\right)^{1/3}\,, & D>0 \\
2\sqrt{-Q}\cos\left(
\frac{1}{3}\cos^{-1}\left(
\frac{P}{\sqrt{-Q^3}}
\right)
\right)
\,, & D<0
\end{array}
\end{align}
with $D \equiv P^2 + Q^3$, $P \equiv \frac{4\lambda_{31}\lambda_{41}\lambda_{412}-8\lambda_{32}\lambda_{41}^2-\lambda_{31}^3}{64\lambda_{41}^3}$, and $Q \equiv \frac{8\lambda_{41}\lambda_{412}-3\lambda^2_{31}}{48\lambda_{41}^2}$.

\section{Dark matter phenomenology}
\label{sec:3}
In this part, we discuss the dark matter phenomenology in the model. We have interactions of dark matter particles with SM particles via the SM Higgs portal and the triplet scalar in the Type-II seesaw. To obtain the observed dark matter relic density, we consider the Freeze-out mechanism.
\subsection{The relic density}
\begin{table}[t]
    
    \begin{tabular}{c |c}
      $\phi_1$ Processes   & Type  \\
      \hline 
        $\phi_1+\phi_1^\dagger\to SM + SM$ & $1100$\\
        $\phi_1+\phi_1^\dagger \to \phi_2+\phi_2^\dagger$  & $1122$\\
        $ \phi_1^\dagger+h \to \phi_2+\phi_2 $  & $1022$\\
        $ \phi_1 + \phi_2^\dagger  \to \phi_2+\phi_2 $ & $1222$\\
        $\phi_1^\dagger + \phi_1^\dagger \to \phi_2+\phi_1  $ & $1112$\\
        $\phi_1+\phi_2 \to \phi_2^\dagger + h$ & $1220$\\
        $\phi_1+\phi_1 \to  \phi_2 + h $ & $1120$\\
        $\phi_1+\phi_1^{\dagger} \to  \Delta + \Delta $ & $1133$\\
    \end{tabular}\hspace{2cm}
    \begin{tabular}{c |c}
      $\phi_2$ Processes   & Type  \\
      \hline 
        $\phi_2+\phi_2^\dagger\to SM + SM$ & $2200$\\
        $\phi_2+\phi_2^\dagger \to \phi_1+\phi_1^\dagger$  & $2211$\\
        $\phi_2+\phi_2 \to \phi_1^\dagger+h$  & $2210$\\
        $\phi_2+\phi_2 \to \phi_1 + \phi_2^\dagger$ & $2212$\\
        $\phi_2+\phi_1 \to \phi_1^\dagger + \phi_1^\dagger$ & $2111$\\
        $\phi_2+\phi_1^\dagger \to \phi_1 + h$ & $2110$\\
        $ \phi_2 + h\to \phi_1+\phi_1 $ & $2011$\\
        $\phi_2+\phi_2^{\dagger} \to  \Delta + \Delta $ & $2233$\\
    \end{tabular}
    
    \caption{The allowed $2\to 2$ processes in the model that can modify the relic density of $\phi_1$ (left) and $\phi_2$ (right). $h$ denotes the SM Higgs boson and $\Delta$ denotes the triplet scalar. Conjugate and inverse processes are not shown.  
    }
    \label{tab1}
\end{table}
The full set of $2 \to 2$ processes that may contribute to the relic density in an arbitrary two-component dark matter scenario in the SM can be found in \cite{Belanger:2014vza}. In this work, we have additional DM-triplet interactions, so that the related interactions can be classified into types that are denoted by four digits (each a 0, 1,2,3) indicating the sector to which the particles involved in the process belong to — 0 is used for SM particles, 1 for $\phi_1$ or $\phi_1^{\dagger}$, 2 for
$\phi_2$ or $\phi_2^{\dagger}$, and 3 for the triplet particles ($\delta^{++(--)},\delta^{+(-)} ,\delta^0,\eta^0$). Therefore, type 1120 includes all processes with one SM particle and one $\phi_2$ (or
 $\phi_2^{\dagger}$) in the final state, and with an initial state consisting of either two $\phi_1$, two $\phi_1^{\dagger}$, or $\phi_1$
and $\phi_1^{\dagger}$. Among the various types, the only ones not compatible with the $Z_5$ symmetry are 1110 and 2220. We display all the processes that contribute to the relic densities in our model in Table.~\ref{tab1}, with their respective type.
\begin{figure}
\centering
\includegraphics[scale=0.9]{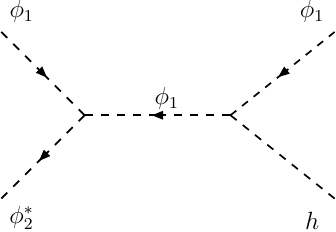}\hspace{0.4cm}
\includegraphics[scale=0.9]{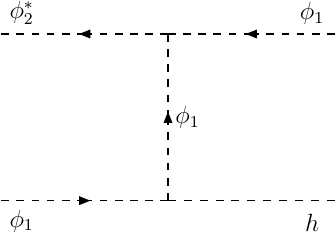}\hspace{0.4cm}
\includegraphics[scale=0.9]{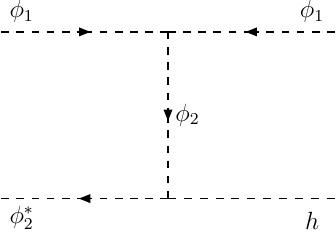}\\
\vspace{0.4cm}
\includegraphics[scale=0.9]{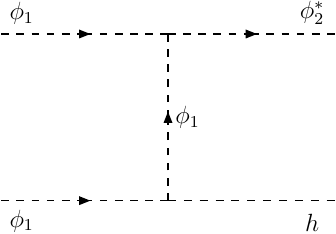}\hspace{1cm}
\includegraphics[scale=0.9]{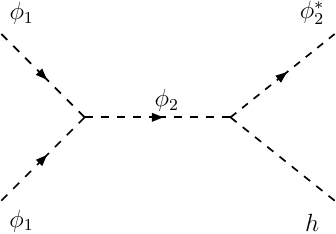}
\caption{Dark matter semi-annihilation processes, figures taken from \cite{Belanger:2020hyh}}
\label{fig1}
\end{figure}
\begin{figure}
\centering
\includegraphics[scale=0.9]{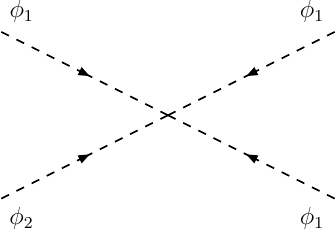}\hspace{0.4cm}
\includegraphics[scale=0.9]{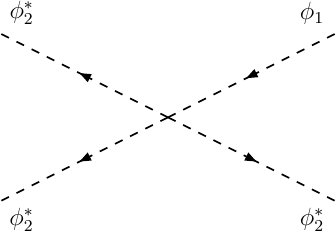}\hspace{0.4cm}
\includegraphics[scale=0.9]{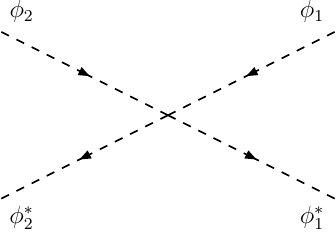}\\
\includegraphics[scale=0.9]{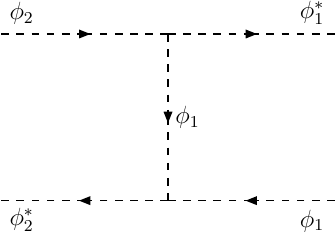}\hspace{1cm}
\includegraphics[scale=0.9]{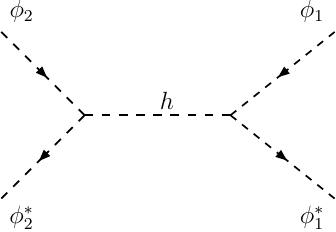}
\caption{Dark matter conversion processes, figures taken from \cite{Belanger:2020hyh}}
\label{fig2}Unable
\end{figure}
\begin{figure}
\centering
\includegraphics[scale=0.65]{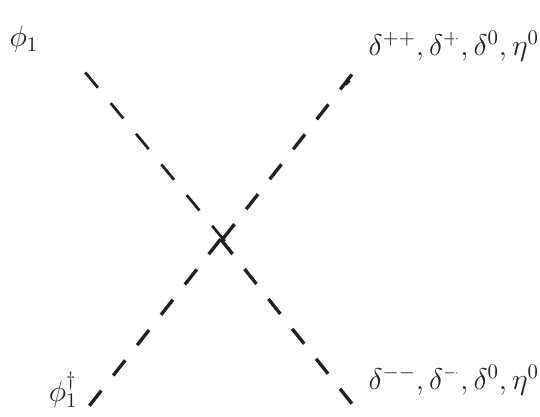}\hspace{0.4cm}
\includegraphics[scale=0.65]{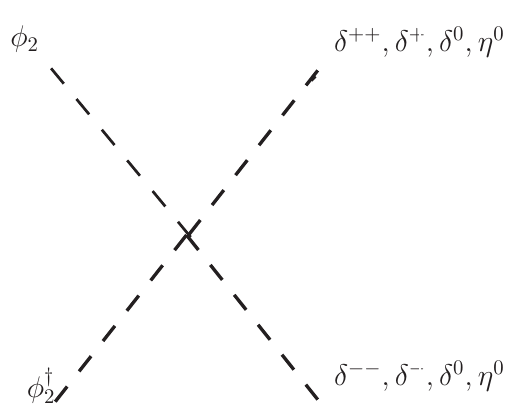}\hspace{0.4cm}
\caption{Dark matter annihilating into triplets.}
\label{fig3}
\end{figure}

 The relevant processes can be divided into three kinds: annihilation processes (including DM pairs annihilating into SM particles and triplet), semi-annihilation processes (including only one SM particle) and dark matter particle conversion processes. In Fig.~\ref{fig1} and Fig.~\ref{fig2}, we give the Feynman diagrams for semi-annihilation and dark matter conversion processes respectively, and in Fig.~\ref{fig3}, we give the processes that dark matter annihilating into triplet particles.

  The Boltzmann equations for the dark matter particles are given as follows:
\begin{eqnarray}
\label{boltzmann1}
\frac{dn_1}{dt} +3H n_1&=&-\sigma_v^{1100}  \left(n_1^2-\bar{n}_1^2 \right) -
\sigma_v^{1120}\left( n_1^2- n_2 \frac{\bar{n}_1^2}{\bar{n}_2} \right)
- \sigma_v^{1122}\left( n_1^2- n_2^2 \frac{\bar{n}_1^2}{\bar{n}_2^2}
\right)\nonumber  \\
  && - \frac{1}{2}\sigma_v^{1112}\left( n_1^2- n_1 n_2 \frac{\bar{n}_1}{\bar{n}_2}\right)
      - \frac{1}{2}\sigma_v^{1222}\left( n_1 n_2- n_2^2\frac{\bar{n}_1}{\bar{n}_2}\right)\nonumber \\
               &&  -\frac{1}{2}\sigma_v^{1220}\left( n_1 n_2- n_2 \bar{n}_1 \right)
+\frac{1}{2}\sigma_v^{2210}(n_2^2-n_1\frac{\bar{n}_2^2}{\bar{n}_1}) -\sigma_v^{1133}\left( n_1^2- \bar{n}_1^2\right)  \,, \\
\frac{dn_2}{dt}  + 3H n_2&=&-\sigma_v^{2200}  \left(n_2^2-\bar{n}_2^2 \right) -
\sigma_v^{2210}\left( n_2^2- n_1 \frac{\bar{n}_2^2}{\bar{n}_1} \right)
- \sigma_v^{2211}\left( n_2^2- n_1^2 \frac{\bar{n}_2^2}{\bar{n}_1^2}
\right)\nonumber  \\
  && - \frac{1}{2}\sigma_v^{2221}\left( n_2^2- n_1 n_2 \frac{\bar{n}_2}{\bar{n}_1}\right)
      - \frac{1}{2}\sigma_v^{1211}\left( n_1 n_2- n_1^2\frac{\bar{n}_2}{\bar{n}_1}\right)\nonumber \\
               &&  -\frac{1}{2}\sigma_v^{1210}\left( n_1 n_2- n_1 \bar{n}_2 \right)
+\frac{1}{2}\sigma_v^{1120}(n_1^2-n_2\frac{\bar{n}_1^2}{\bar{n}_2}) -\sigma_v^{2233}\left( n_2^2- \bar{n}_2^2\right)  .      
\label{boltzmann2}
\end{eqnarray}
Here $n_{i}$ ($i=1,2$) denote the number densities of $\phi_i$, and $\bar{n}_i$  their respective equilibrium values.  $\sigma_v^{abcd}$  is  the thermally averaged cross section, which satisfies: 
\begin{equation}
    \bar{n}_a\bar{n}_b\sigma_v^{abcd}=\bar{n}_c\bar{n}_d\sigma_v^{cdab}.
\end{equation}

 By solving the above Boltzmann equations, we can obtain the relic densities of dark matter particles $\phi_1$ and $\phi_2$. Numerically, we calculate the relic densities with the micrOMEGAs package \cite{Belanger:2018ccd}, and the implemented model file is generated by Feynrules \cite{Dedes:2023zws}.

\subsection{Parameter dependence}
\begin{figure}[htbp]
\centering
\subfigure[]{\includegraphics[height=4.9cm,width=4.9cm]{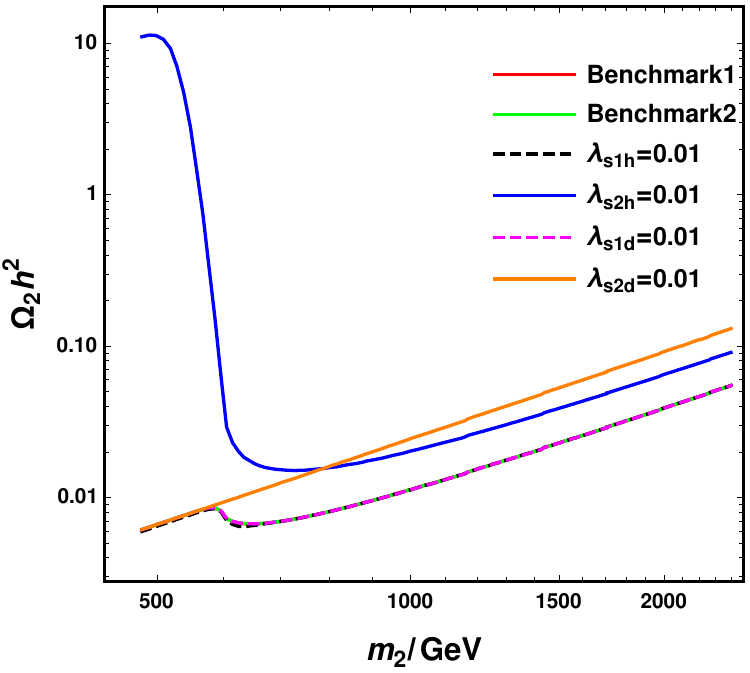}}
\subfigure[]{\includegraphics[height=4.9cm,width=4.9cm]{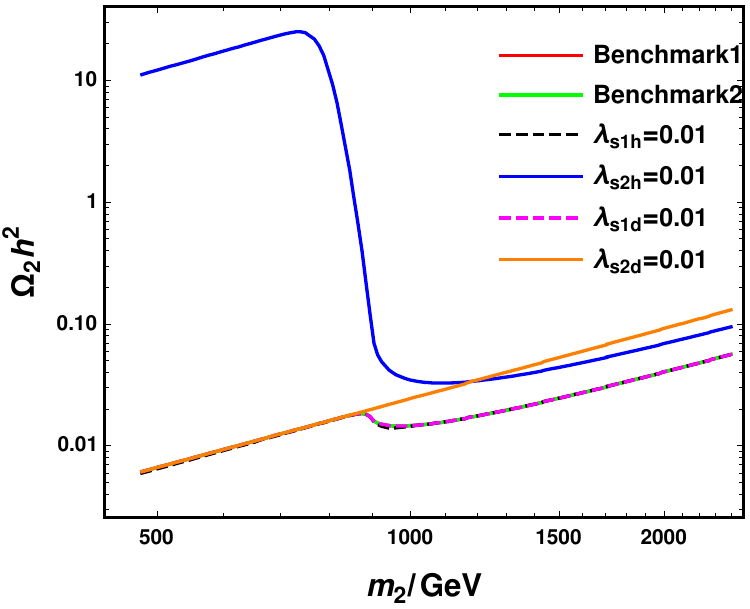}}
\subfigure[]{\includegraphics[height=4.9cm,width=4.9cm]{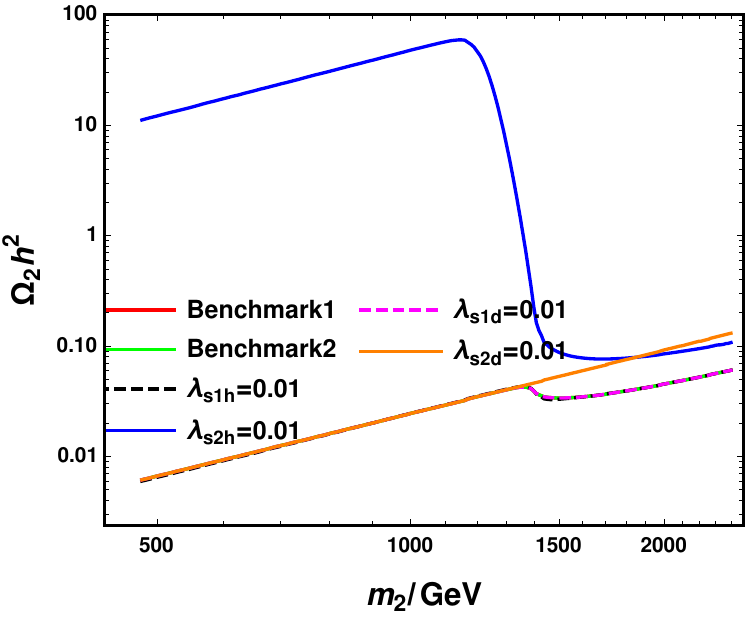}}
\caption{Evolution of $\Omega_2h^2$ with $m_2$, where $m_2=1.2m_1$, $M_{\Delta}= 600,900,1400$ GeV. The red lines in each picture correspond to the case of  $\lambda_{412}=\mu_{si}=\lambda_{3i}=0$, $\lambda_{sid}=\lambda_{sih}=1$ with $i=1,2$, the green dashed lines in each picture correspond to the case of  $\lambda_{412}=\lambda_{3i}=0.01$, $\mu_{si}=0.01$ GeV, $\lambda_{sid}=\lambda_{sih}=1$ with $i=1,2$, and other colored lines represent the case that we vary one of the chosen parameters.  }
\label{fig4}
\end{figure}
\begin{figure}[htbp]
\centering
\subfigure[]{\includegraphics[height=4.9cm,width=4.9cm]{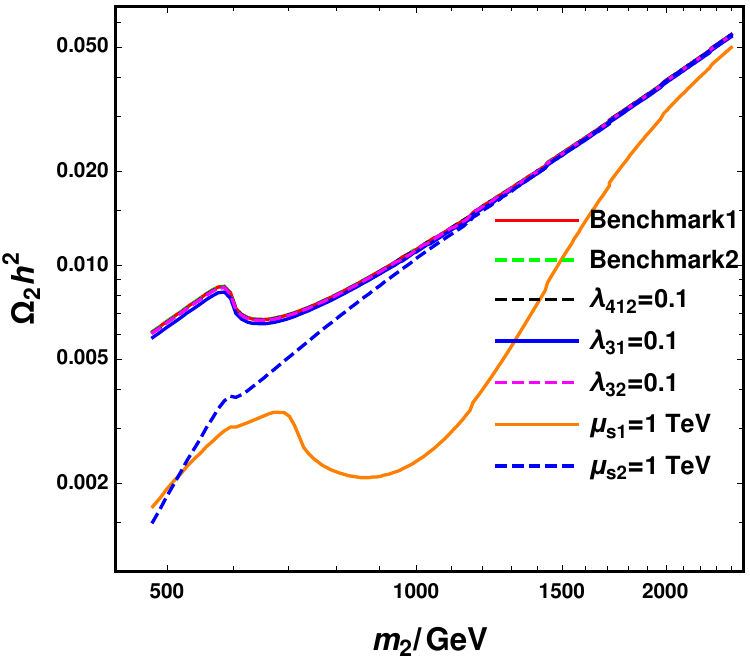}}
\subfigure[]{\includegraphics[height=4.9cm,width=4.9cm]{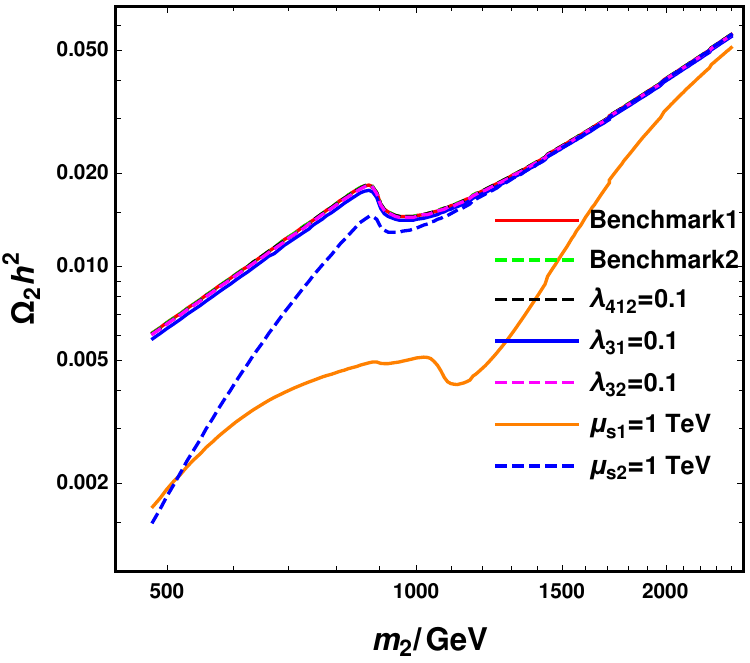}}
\subfigure[]{\includegraphics[height=4.9cm,width=4.9cm]{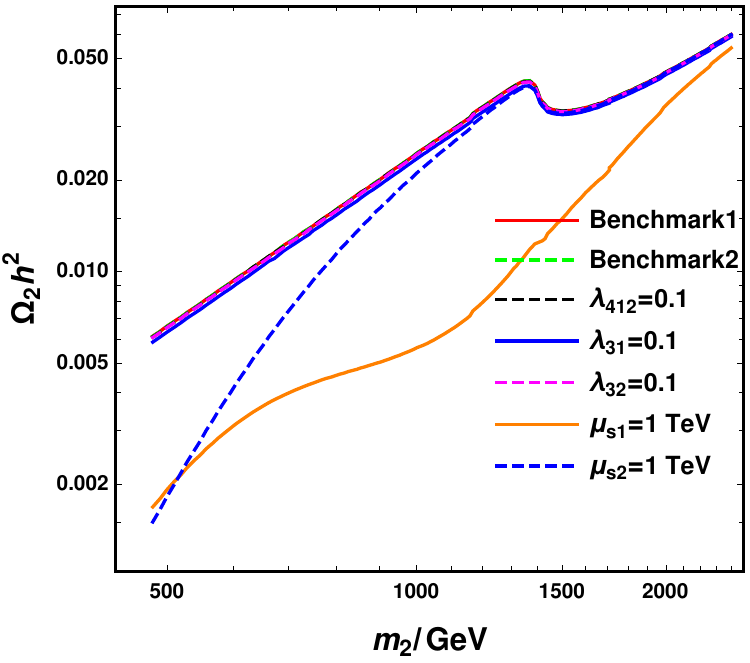}}
\caption{Evolution of $\Omega_2h^2$ with $m_2$, where $m_2=1.2m_1$, $M_{\Delta}= 600,900,1400$ GeV. The red lines in each picture correspond to the case of  $\lambda_{412}=\mu_{si}=\lambda_{3i}=0$, $\lambda_{sid}=\lambda_{sih}=1$ with $i=1,2$, the green dashed lines in each picture correspond to the case of  $\lambda_{412}=\lambda_{3i}=0.01$,$\mu_{si}=0.01$ GeV, $\lambda_{sid}=\lambda_{sih}=1$ with $i=1,2$, and other colored lines represent the case that we vary one of the chosen parameters. }
\label{fig5}
\end{figure}
\begin{figure}[htbp]
\centering
\subfigure[]{\includegraphics[height=4.9cm,width=4.9cm]{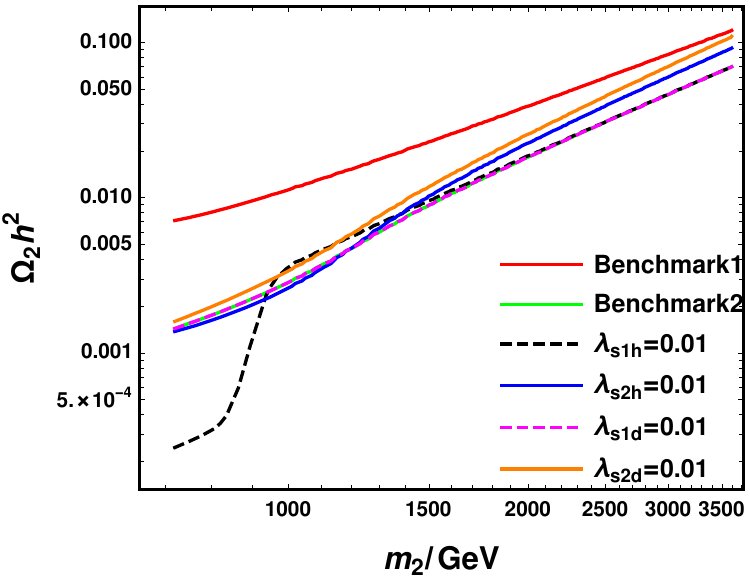}}
\subfigure[]{\includegraphics[height=4.9cm,width=4.9cm]{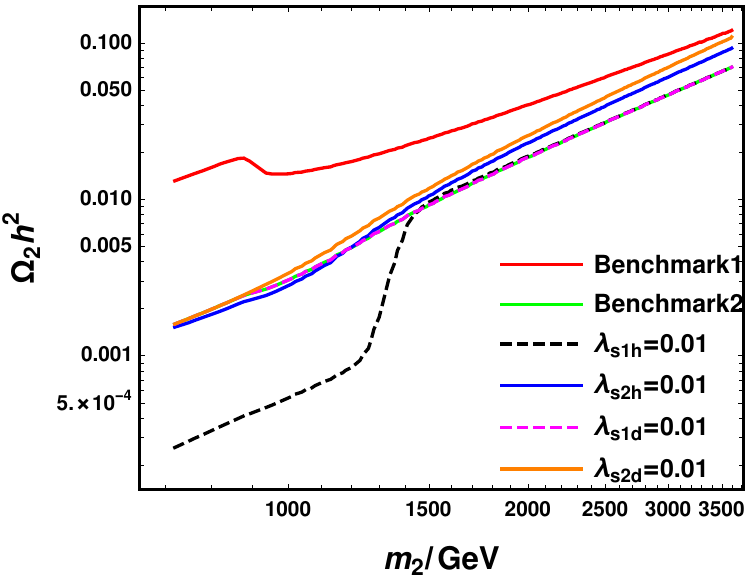}}
\subfigure[]{\includegraphics[height=4.9cm,width=4.9cm]{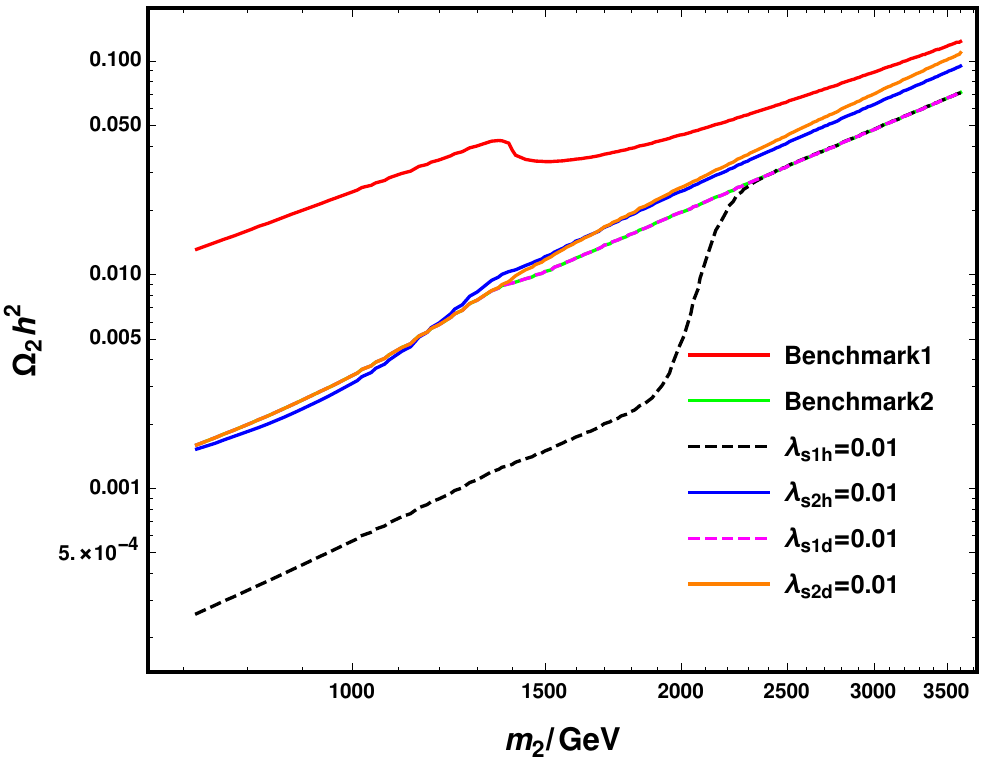}}
\caption{Evolution of $\Omega_2h^2$ with $m_2$, where $m_2=1.8m_1$, $M_{\Delta}= 600,900,1400$ GeV. The red lines in each picture correspond to the case of  $\lambda_{412}=\mu_{si}=\lambda_{3i}=0$, $\lambda_{sid}=\lambda_{sih}=1$ with $i=1,2$, the green dashed lines in each picture correspond to the case of  $\lambda_{412}=\lambda_{3i}=0.01$, $\mu_{si}=0.01$ GeV, $\lambda_{sid}=\lambda_{sih}=1$ with $i=1,2$, and other colored lines represent the case that we vary one of the chosen parameters. }
\label{fig6}
\end{figure}
\begin{figure}[htbp]
\centering
\subfigure[]{\includegraphics[height=4.9cm,width=4.9cm]{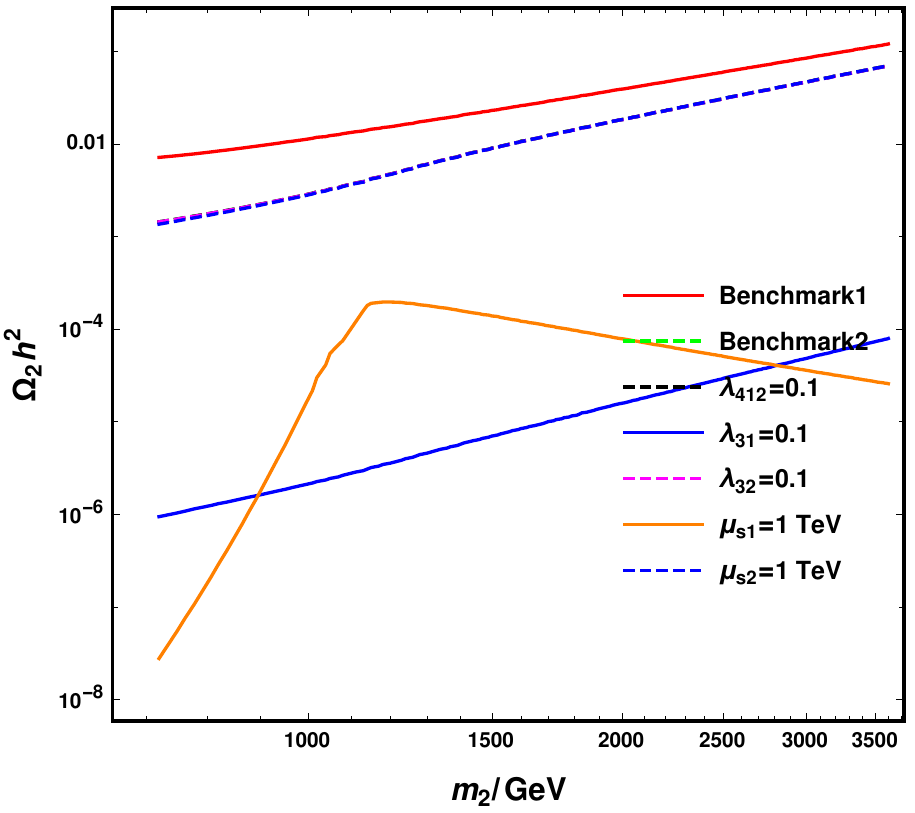}}
\subfigure[]{\includegraphics[height=4.9cm,width=4.9cm]{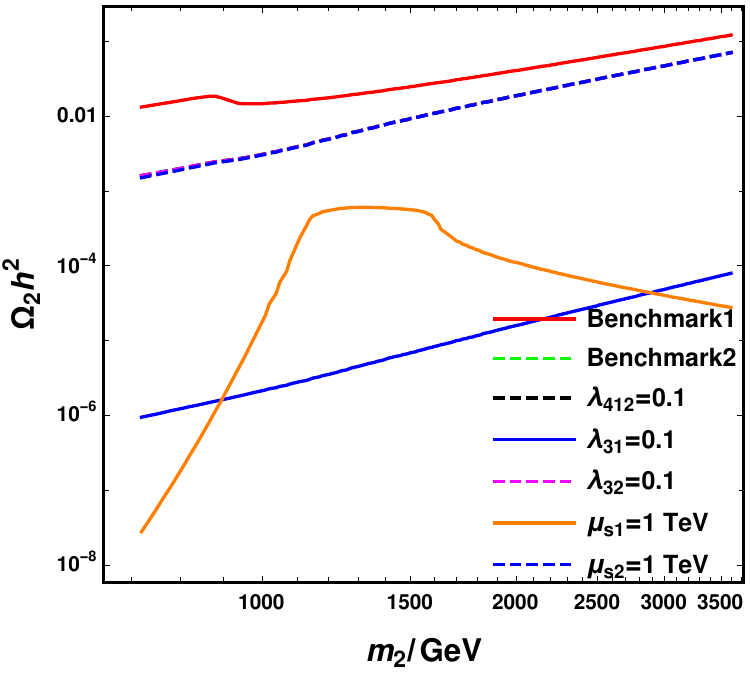}}
\subfigure[]{\includegraphics[height=4.9cm,width=4.9cm]{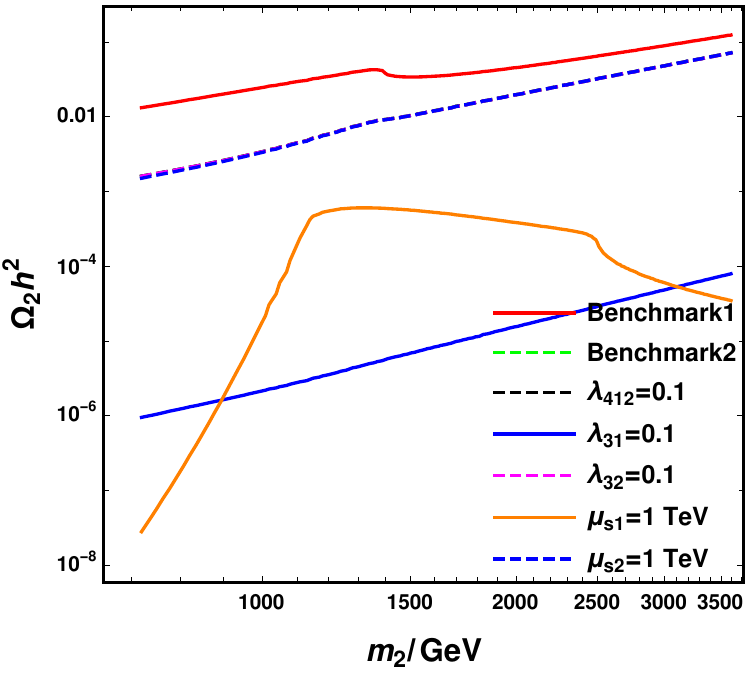}}
\caption{Evolution of $\Omega_2h^2$ with $m_2$, where $m_2=1.8m_1$, $M_{\Delta}= 600,900,1400$ GeV. The red lines in each picture correspond to the case of  $\lambda_{412}=\mu_{si}=\lambda_{3i}=0$, $\lambda_{sid}=\lambda_{sih}=1$ with $i=1,2$, the green dashed lines in each picture correspond to the case of  $\lambda_{412}=\lambda_{3i}=0.01$, $\mu_{si}=0.01$ GeV, $\lambda_{sid}=\lambda_{sih}=1$ with $i=1,2$, and other colored lines represent the case that we vary one of the chosen parameters. }
\label{fig7}
\end{figure}
The related parameters to dark matter are given as follows:
\begin{eqnarray*}
    m_i,\lambda_{sih},\lambda_{sid},\mu_{si},\lambda_{3i}, \lambda_{4i},\lambda_{412},M_{\Delta}   ~~~~(i=1,2)
\end{eqnarray*}
Note that $\lambda_{41}$ and $\lambda_{42}$ are irrelevant for the dark matter phenomenology and can be ignored in our analysis. There are 12 free parameters as inputs in our model related to dark matter, and one can divide these parameters into three kinds: mass terms ($m_{1,2}, M_{\Delta}$), couplings related to $Z_5$ two-component dark matter model ($\mu_{s1},\mu_{s2},\lambda_{412}$,$\lambda_{31},\lambda_{32}$) and couplings related to dark matter with scalars ($\lambda_{s1h},\lambda_{s2h},\lambda_{s1d},\lambda_{s2d}$). On one hand, so many parameters can lower the model's predictability. On the other hand, one can have new annihilation channels in the model to have a wider parameter space satisfied with experiment constraints. The phases of $\phi_{1,2}$ can be chosen to make $\mu_{s1}$ and $\mu_{s2}$ real, but then $\lambda_{31}$ and $\lambda_{32}$ may be complex. In the following, we will stick to real parameters for simplicity.

To discuss the effect of the parameters on the relic densities, we consider two benchmark cases: i) $\mu_{si}$ and $\lambda_{3i}$ are set to be 0 with $i=1,2$, which means $\phi_1$ and $\phi_2$ are only connected by the interaction $|\phi_1|^2|\phi_2|^2$
 and the scalar potential with $Z_5$ symmetry is indistinguishable from that $Z_2 \times Z'_2$ symmetry. ii) fixing the parameters above to be certain non-zero values. From Fig.~\ref{fig4} to Fig.~\ref{fig7}, we give the evolution of $\Omega_2h^2$ with $m_2$, where the green dashed lines correspond to the benchmark points that  $\lambda_{412}=\lambda_{3i}=0.01$, $\mu_{si}=0.01$ GeV, $\lambda_{sid}=\lambda_{sih}=1$ with $i=1,2$, while the red lines correspond to the case that $\mu_{si}=\lambda_{3i} =0 ~(i=1,2)$  with other parameters unchanged. Other colored lines correspond to the results that we vary one of the chosen parameters. In Fig.~\ref{fig4} and Fig.~\ref{fig5}, we give the cases of $m_2=1.2m_1$, where the three pictures in each figure correspond to $M_{\Delta}= 600, 900$ and $1400$ GeV respectively. In both figures, the two benchmark lines almost coincide for the small $\lambda_{Si}$ and $\mu_{si}$ with $i=1,2$. A sharp drop can be found in the region of $M_{\Delta} \approx m_2$, which corresponds to the case that the t-channel processes are opened so that dark matter density decreases quickly. For the smaller $\lambda_{s2d}$ and $\lambda_{s2h}$, the relic density of $\phi_2$ is larger, and such results are consistent with the Freeze-out mechanism. On the other hand, the lines representing $\lambda_{s1h}$ and $\lambda_{s1d}$ almost coincide with the benchmark lines since $\phi_2$ density is less relevant with $\lambda_{s1h}$ and $\lambda_{s1d}$. In Fig.~\ref{fig5}, we give the results of other parameters. Relic density of $\phi_2$ is significantly influenced by the dimensional parameters $\mu_{s1}$ and $\mu_{s2}$, where the larger $\mu_{s1}$ and $\mu_{s2}$ can induce smaller density of $\phi_2$ since most of $\phi_2$ particles are converted into $\phi_2$ via the conversion processes as well as semi-annihilation processes.
 
 In Fig.~\ref{fig6} and Fig.~\ref{fig7}, we give the results of $m_2=1.8m_1$ where the difference between $m_2$ and $m_1$ is relatively large, and the three pictures in each figure correspond to $M_{\Delta}= 600, 900$ and $1400$ GeV respectively. Unlike the case of $m_2=1.2m_1$, the two benchmark cases are different. Owing to the non-zero couplings related to the $Z_5$ symmetry, the conversion processes between $\phi_1$ and $\phi_2$ are efficient so that some of $\phi_2$ particles have been converted into $\phi_1$ and the relic density of $\phi_2$ is smaller compared with the results of none $Z_5$ related couplings. We have different behaviors of $\Omega_2h^2$ related to $m_2$ for the large and small differences between $\phi_1$ and $\phi_2$ according to Fig.~\ref{fig4}(Fig.~\ref{fig5}) and Fig.~\ref{fig6}(Fig.~\ref{fig7}) since processes such as $\sigma_v^{1211}$ are related to $m_2/m_1$.
 
 Note that in this work, we have new annihilation processes related to the Type-II seesaw mechanism. Still, the interactions with $Z_5$ symmetry are identical to the two-component $Z_5$ dark matter in the SM, therefore, the effect of these couplings related to $Z_5$ symmetry to dark matter relic density is similar to the two-component $Z_5$ dark matter in the SM. More details about the effect of these couplings can be found in \cite{Belanger:2020hyh}. However, the parameter space should be limited when considering dark matter relic density constraints, and the viable parameter space will be different due to the new annihilation processes.
 
\subsection{The viable parameter space}
\begin{figure}[htbp]
\centering
\subfigure[]{\includegraphics[height=5.5cm,width=5.5cm]{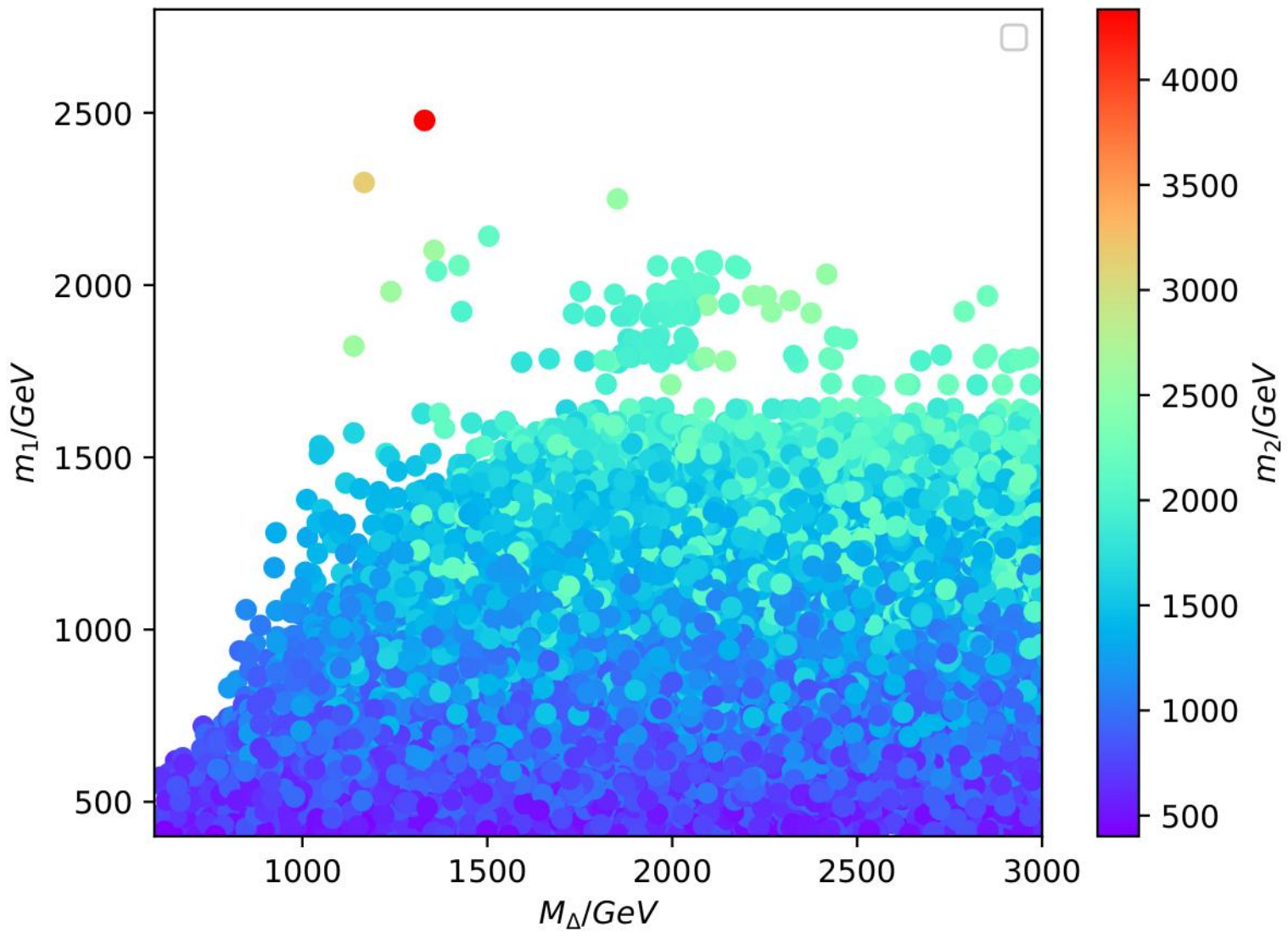}}
\subfigure[]{\includegraphics[height=5.5cm,width=5.5cm]{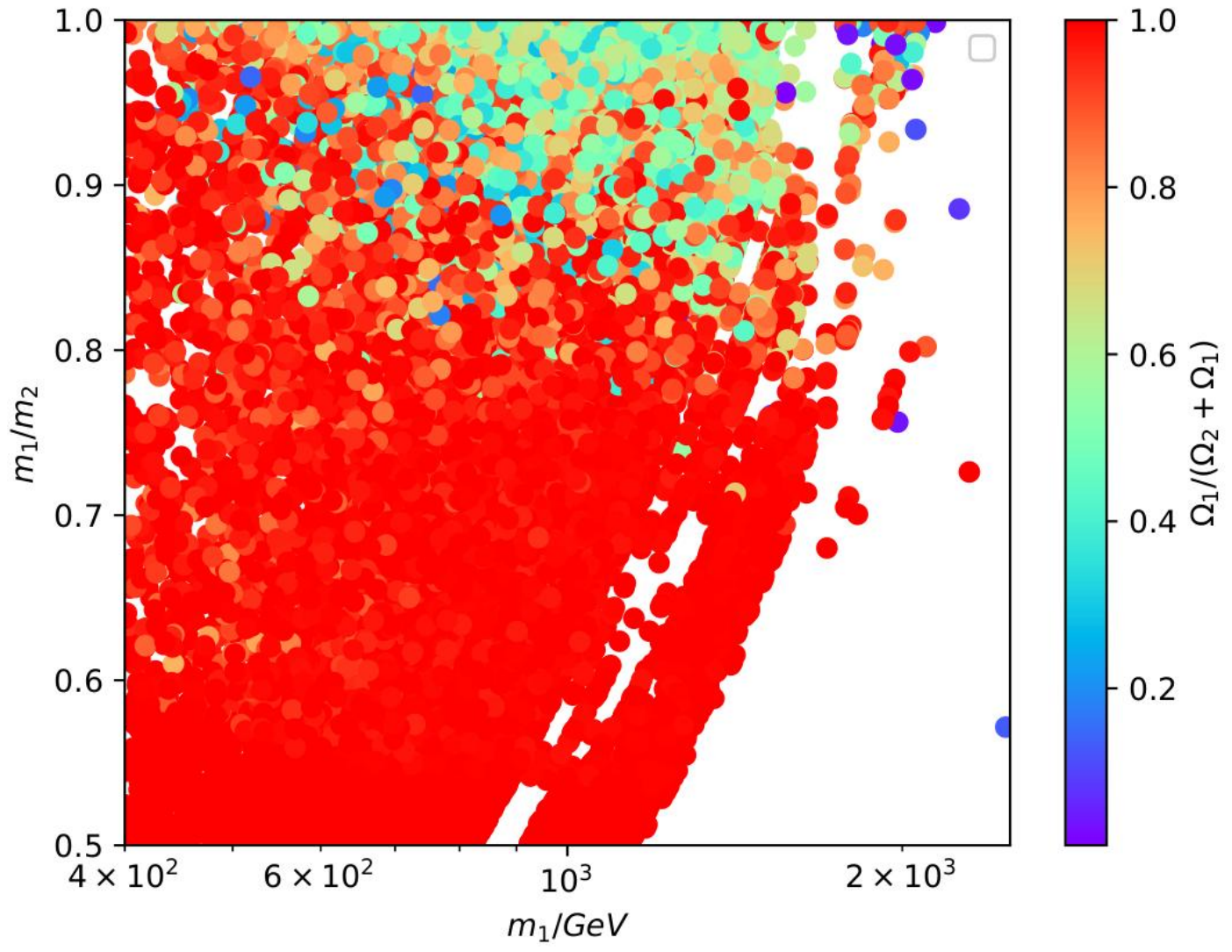}}
\caption{Parameter space satisfying relic density constraint. The picture (a) shows the viable range of $M_{\Delta}-m_1$, where points with different colors correspond $m_2$ taking different values. The picture (b) shows the parameter space of $m_1-m_1/m_2$, where points with different colors represent $\Omega_1/(\Omega_1 +\Omega_2)$. }
\label{fig8}
\end{figure}
\begin{figure}[htbp]
\centering
\subfigure[]{\includegraphics[height=4.9cm,width=4.9cm]{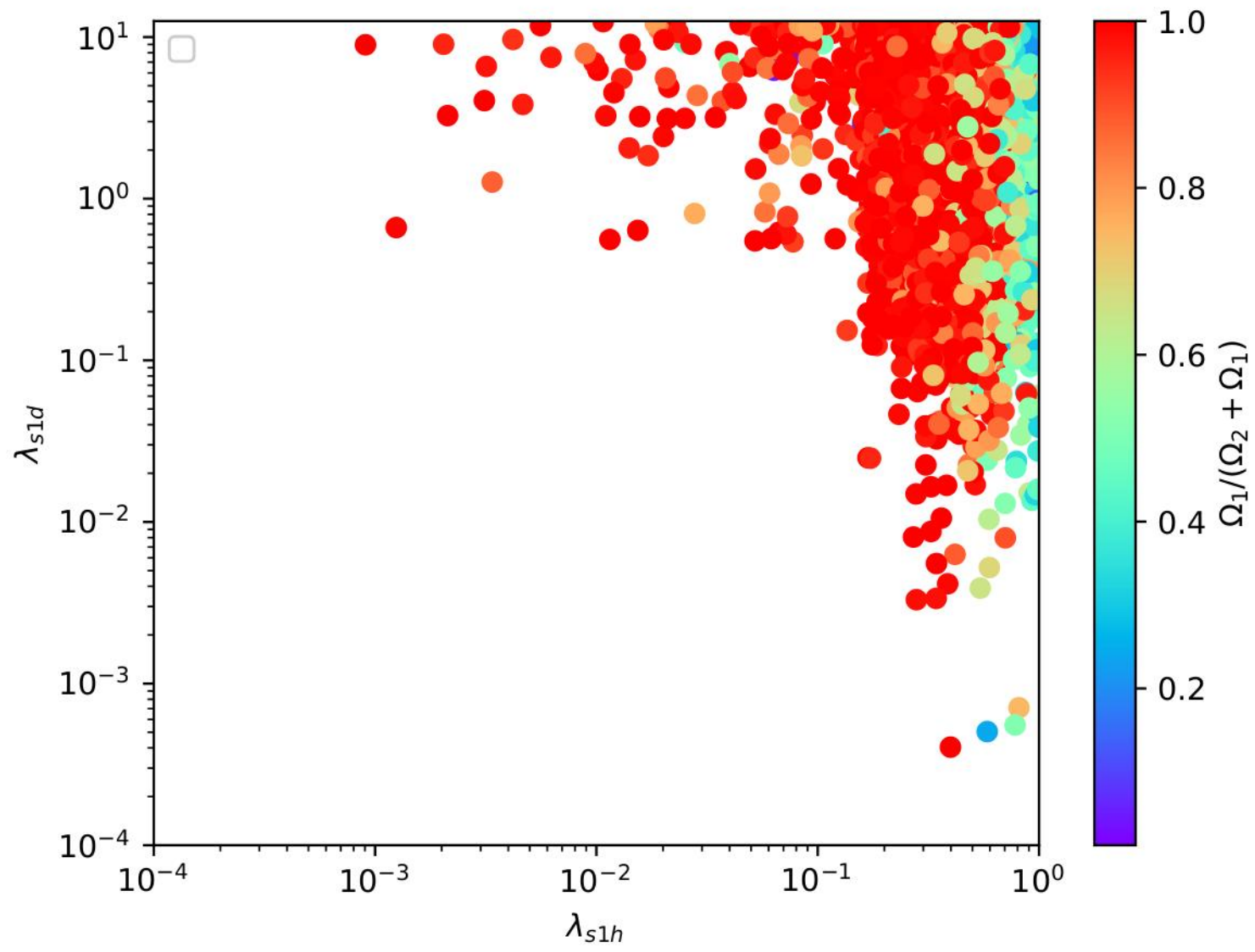}}
\subfigure[]{\includegraphics[height=4.9cm,width=4.9cm]{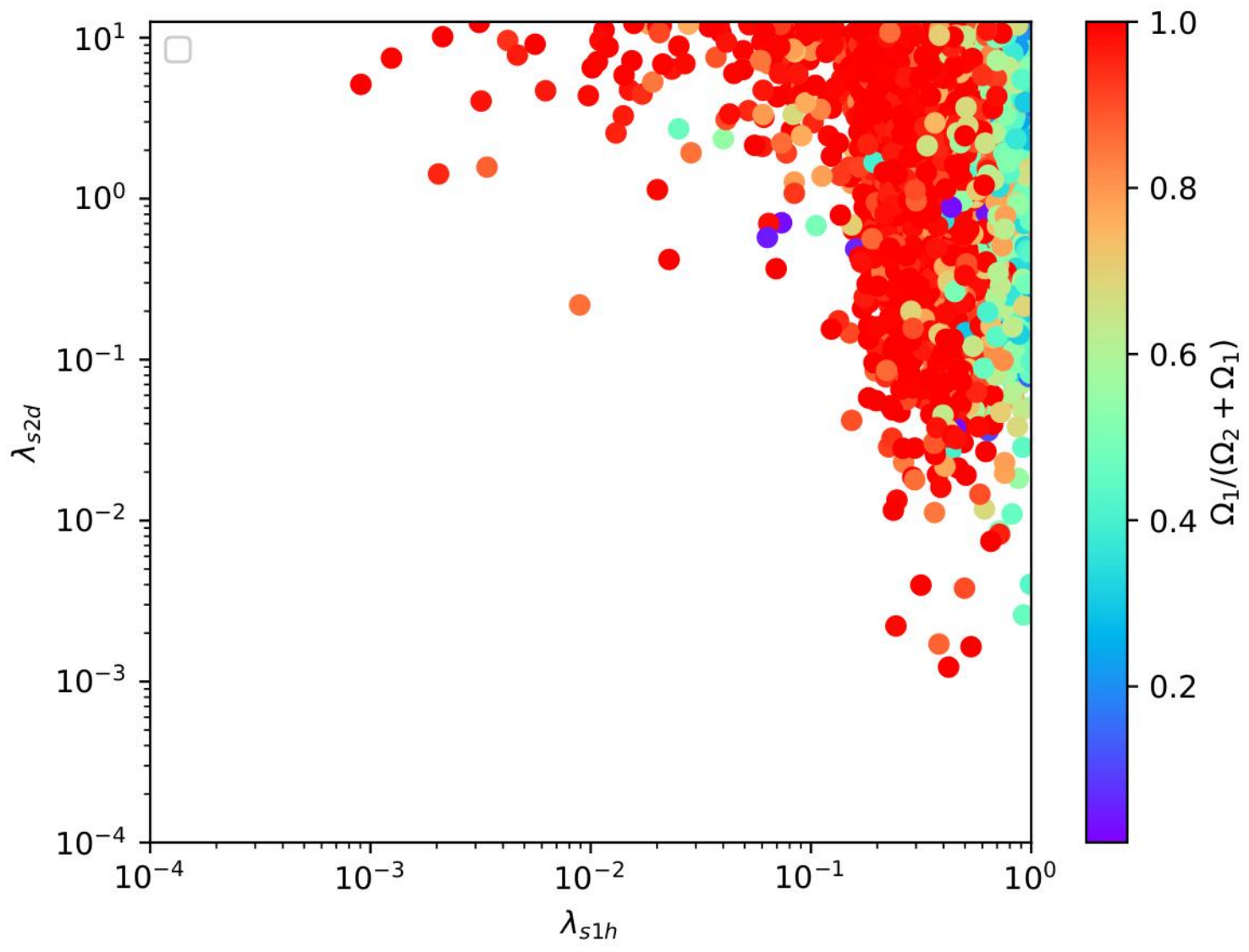}}
\subfigure[]{\includegraphics[height=4.9cm,width=4.9cm]{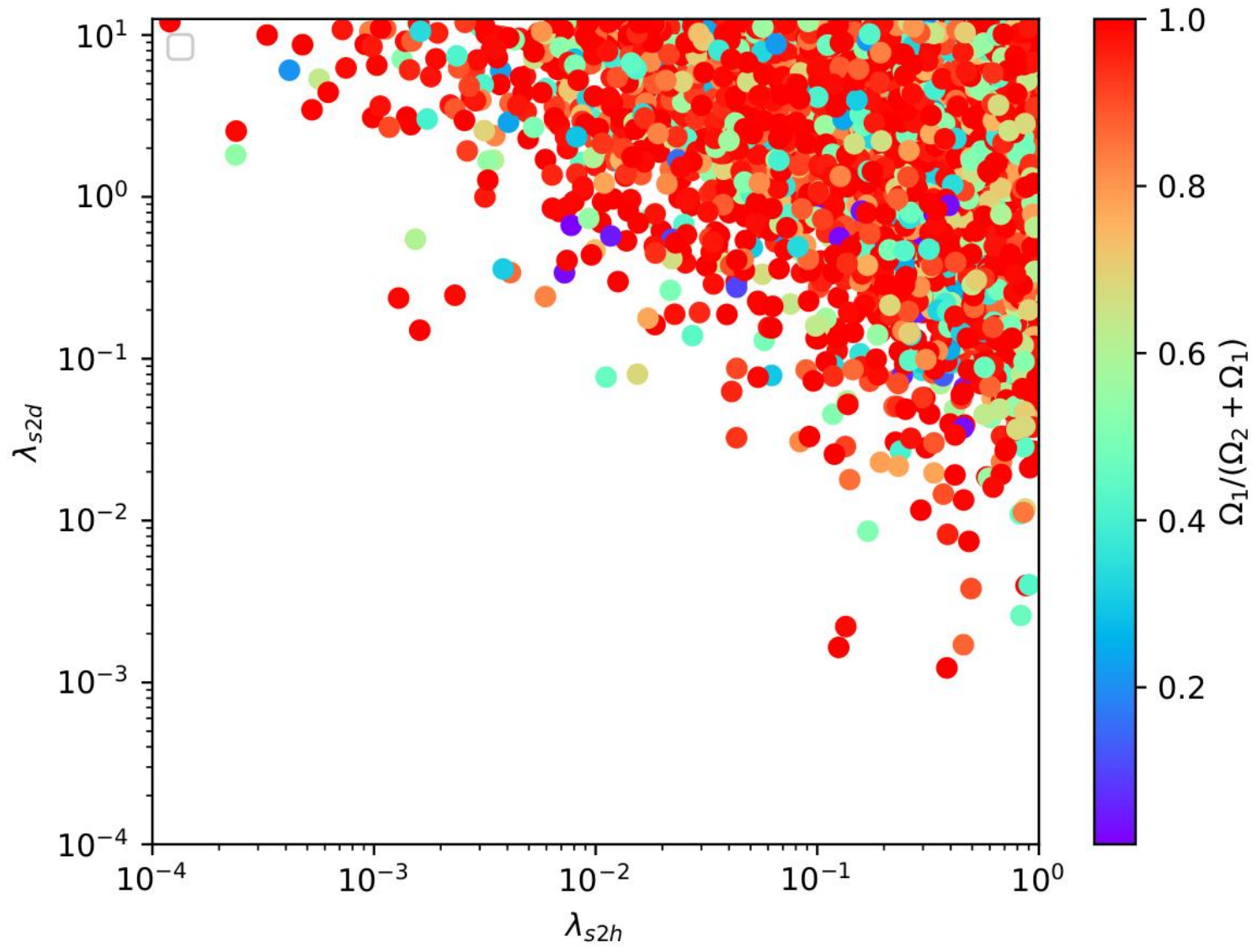}}
\caption{Parameter space satisfying relic density constraint, where points with different colors correspond to the fraction $\Omega_1/(\Omega_1+\Omega_2)$. Three pictures (a),(b) and (c) show the viable region of $\lambda_{s1h}-\lambda_{s1d}$, $\lambda_{s1h}-\lambda_{s2d}$ as well as $\lambda_{s2h}-\lambda_{s2d}$ respectively.}
\label{fig9}
\end{figure}
\begin{figure}[htbp]
\centering
\subfigure[]{\includegraphics[height=5.5cm,width=6.5cm]{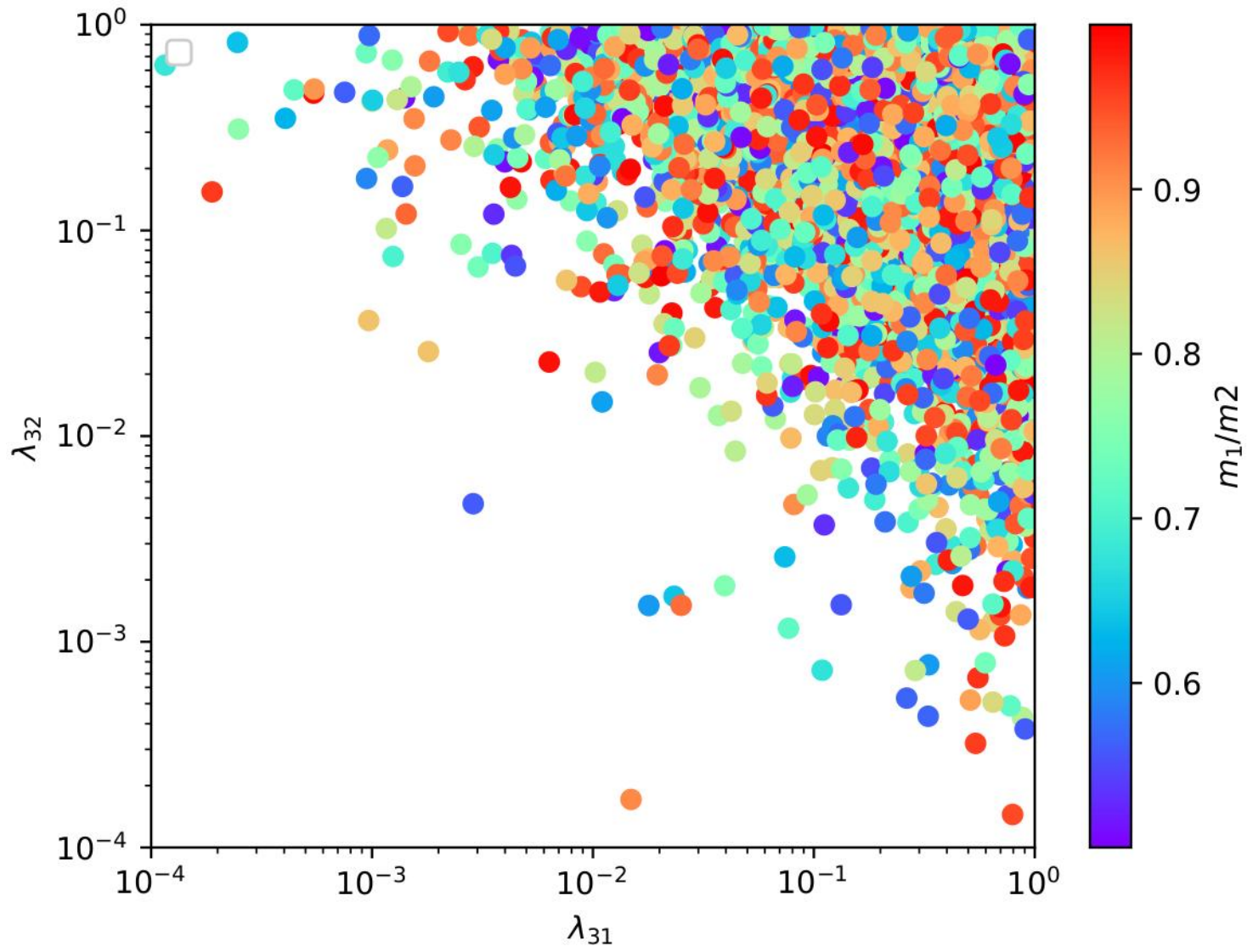}}
\subfigure[]{\includegraphics[height=5.5cm,width=6.5cm]{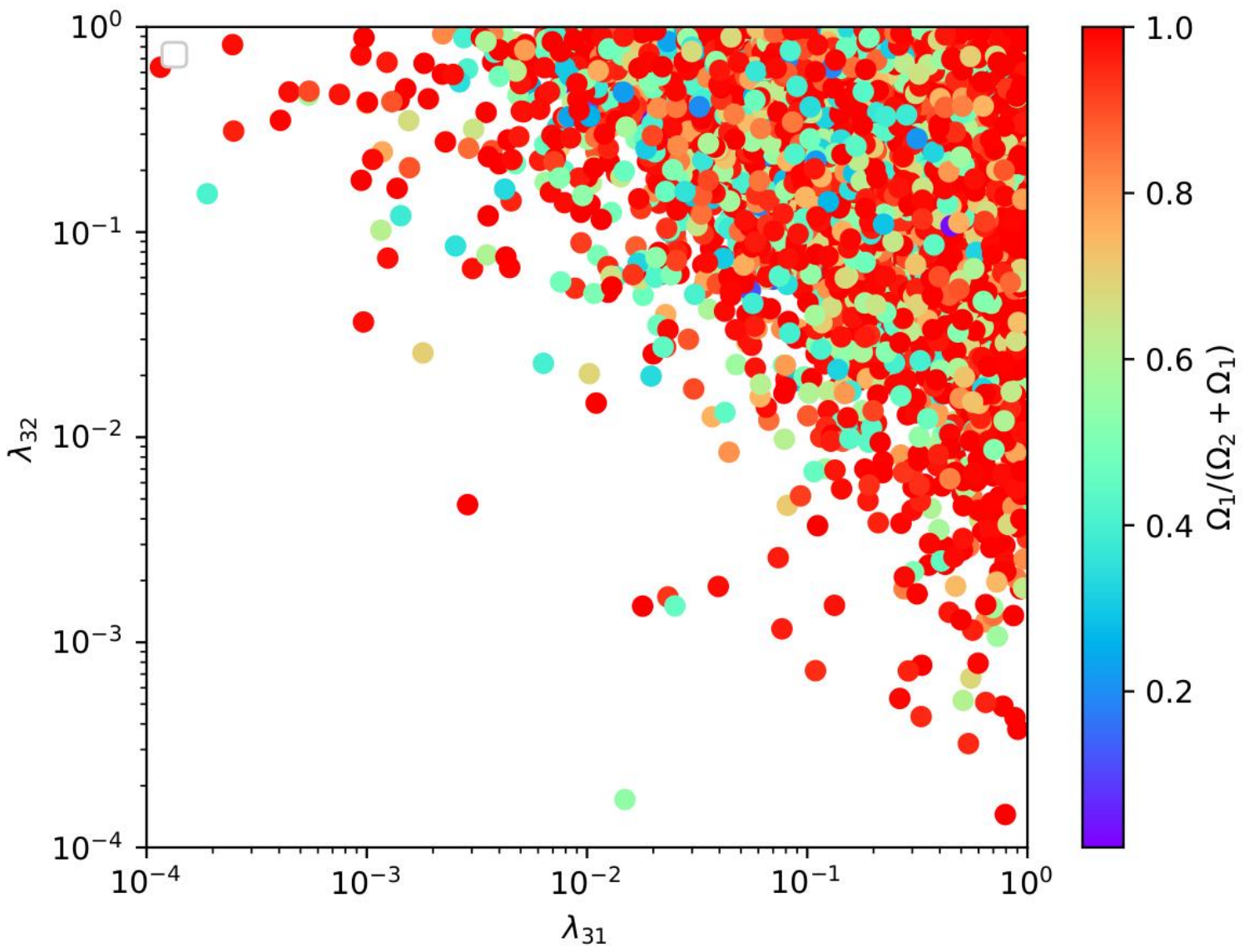}}
\subfigure[]{\includegraphics[height=5.5cm,width=6.5cm]{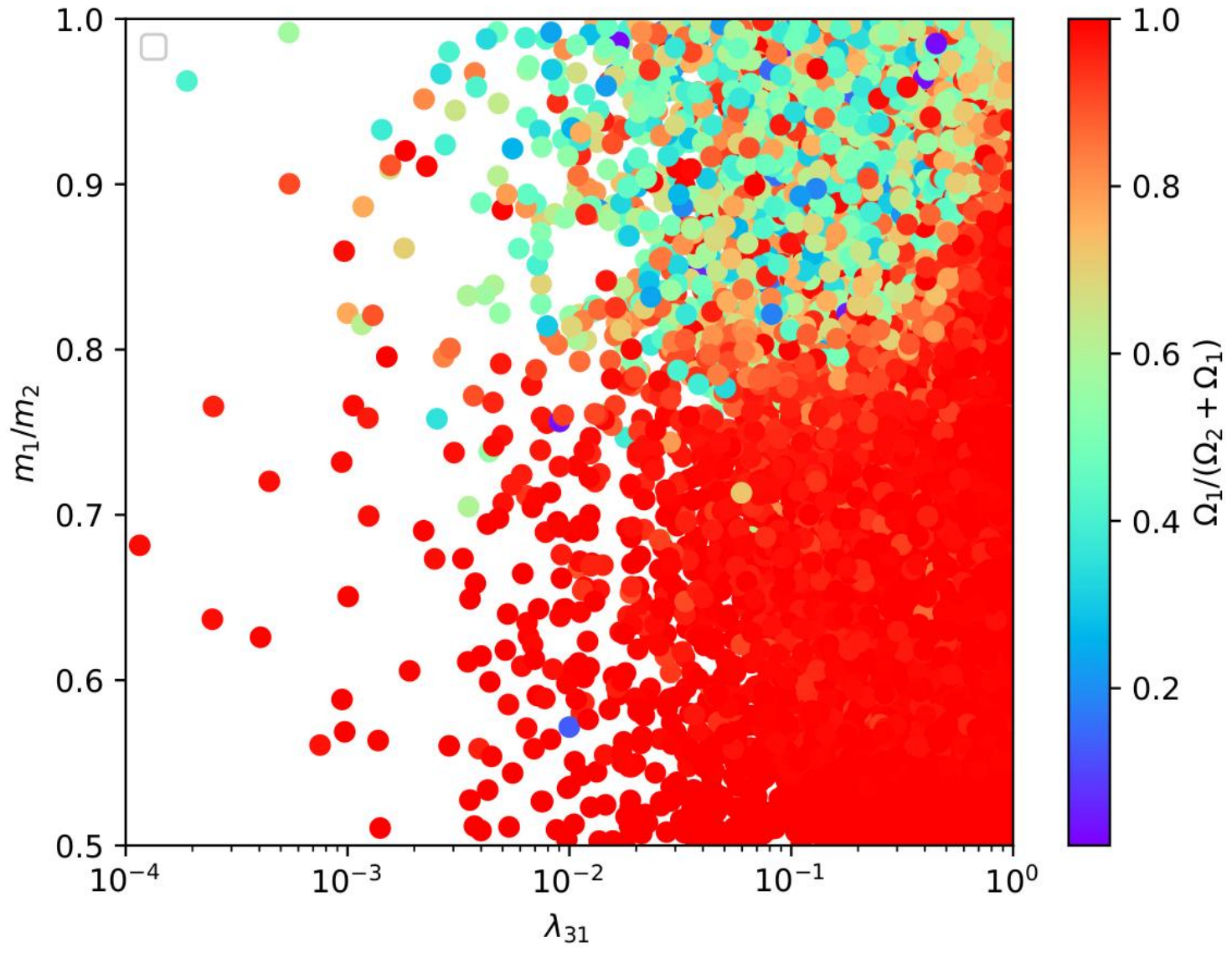}}
\subfigure[]{\includegraphics[height=5.5cm,width=6.5cm]{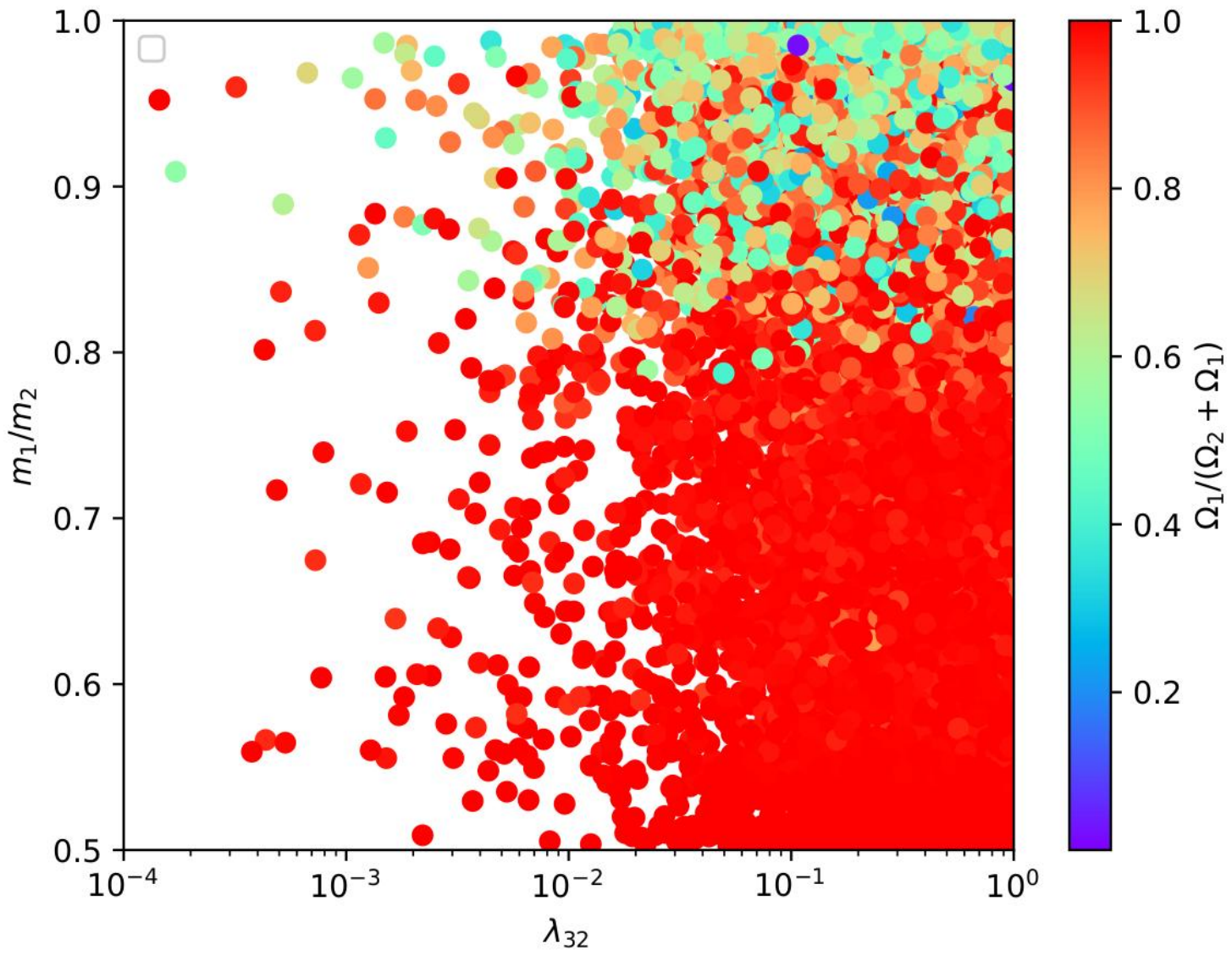}}
\caption{Parameter space satisfying relic density constraint, where points with different colors correspond to $m_1/m_2$ in the picture (a) and the fraction $\Omega_1/(\Omega_1+\Omega_2)$ in the picture (b),(c) and (d).  (a) and (b) show the viable region of $\lambda_{31}-\lambda_{32}$, (c) and (d) show the viable region of $\lambda_{31}-m_1/m_2$ as well as $\lambda_{32}-m_1/m_2$ respectively.}
\label{fig10}
\end{figure}
\begin{figure}[htbp]
\centering
\subfigure[]{\includegraphics[height=5.5cm,width=6.5cm]{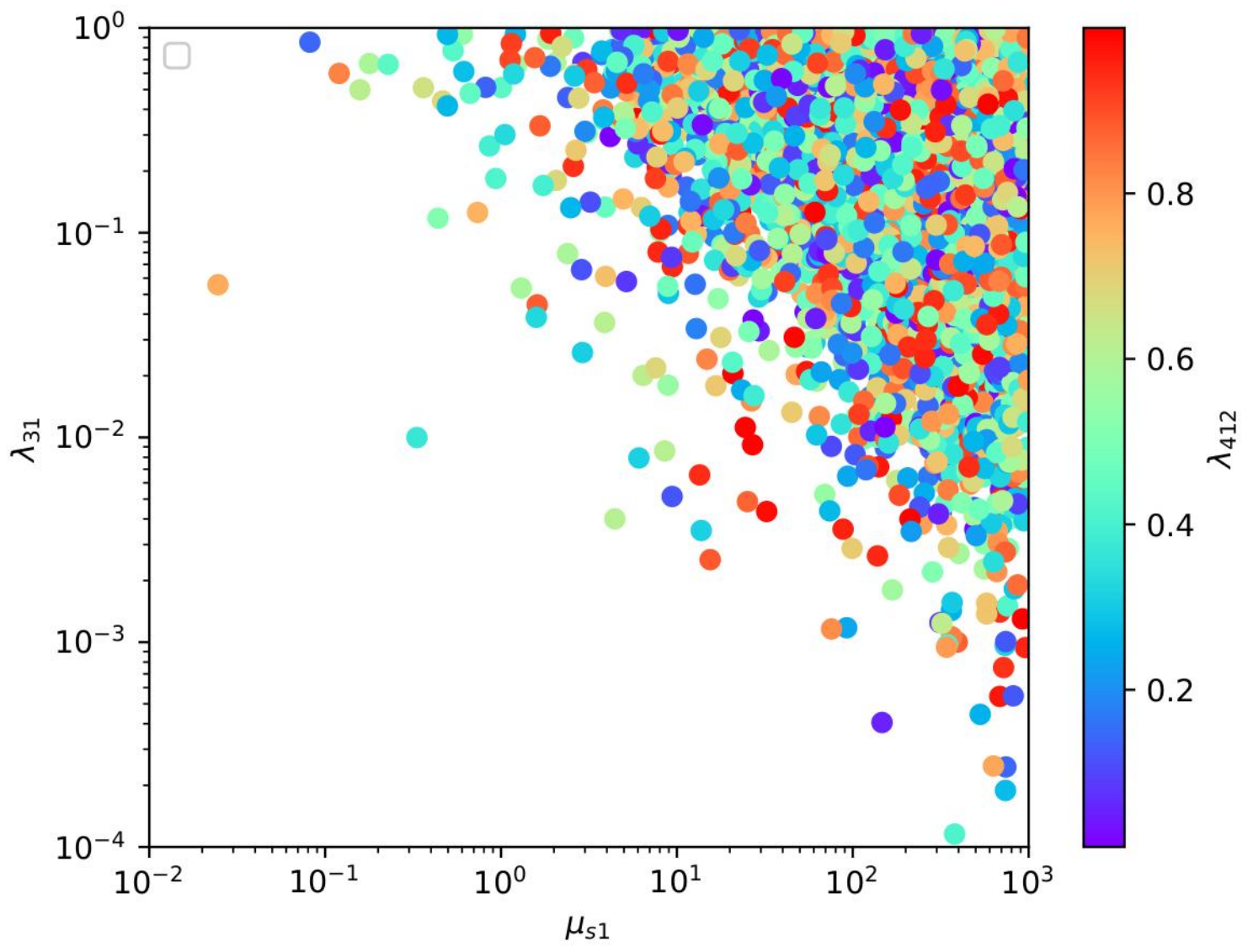}}
\subfigure[]{\includegraphics[height=5.5cm,width=6.5cm]{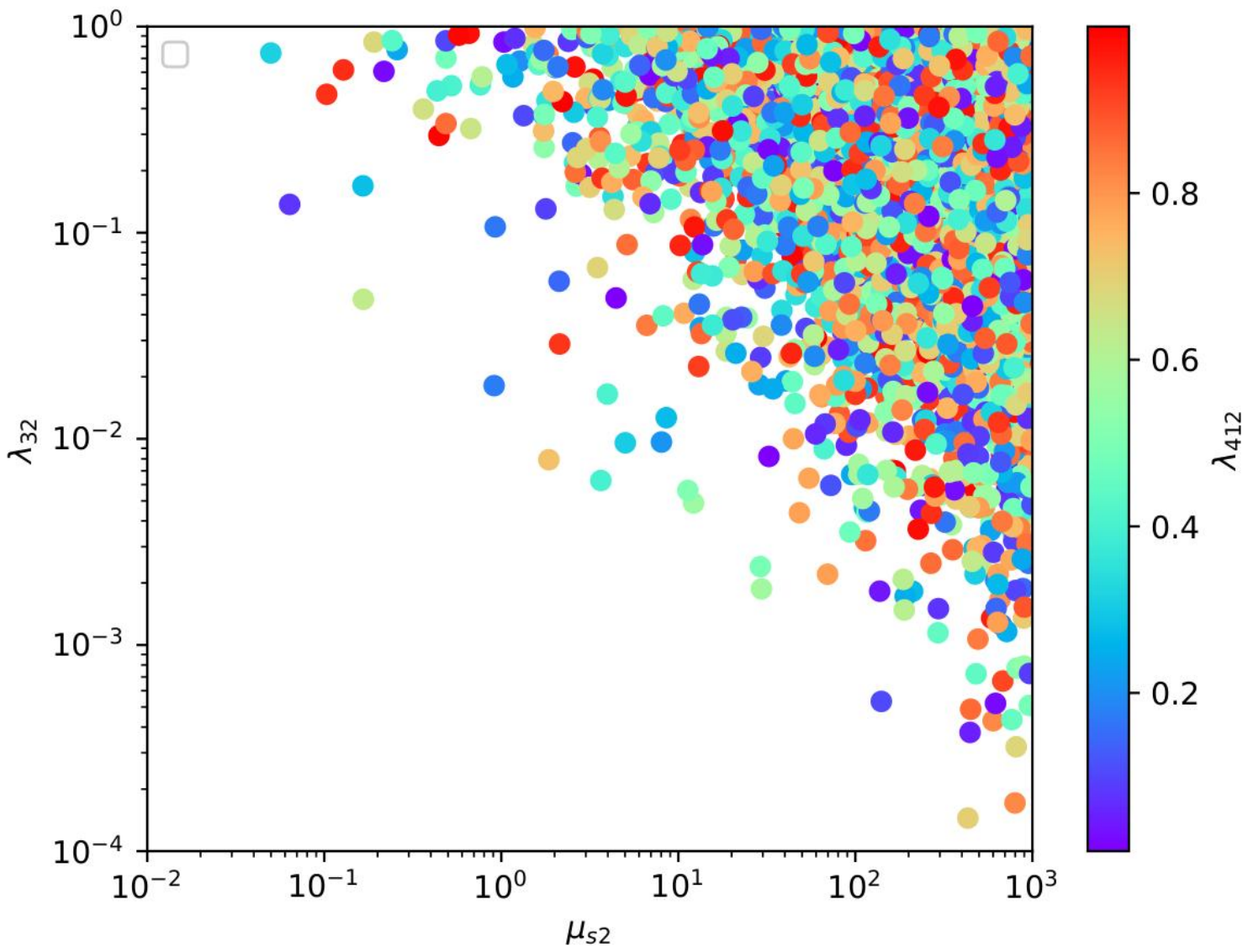}}
\subfigure[]{\includegraphics[height=5.5cm,width=6.5cm]{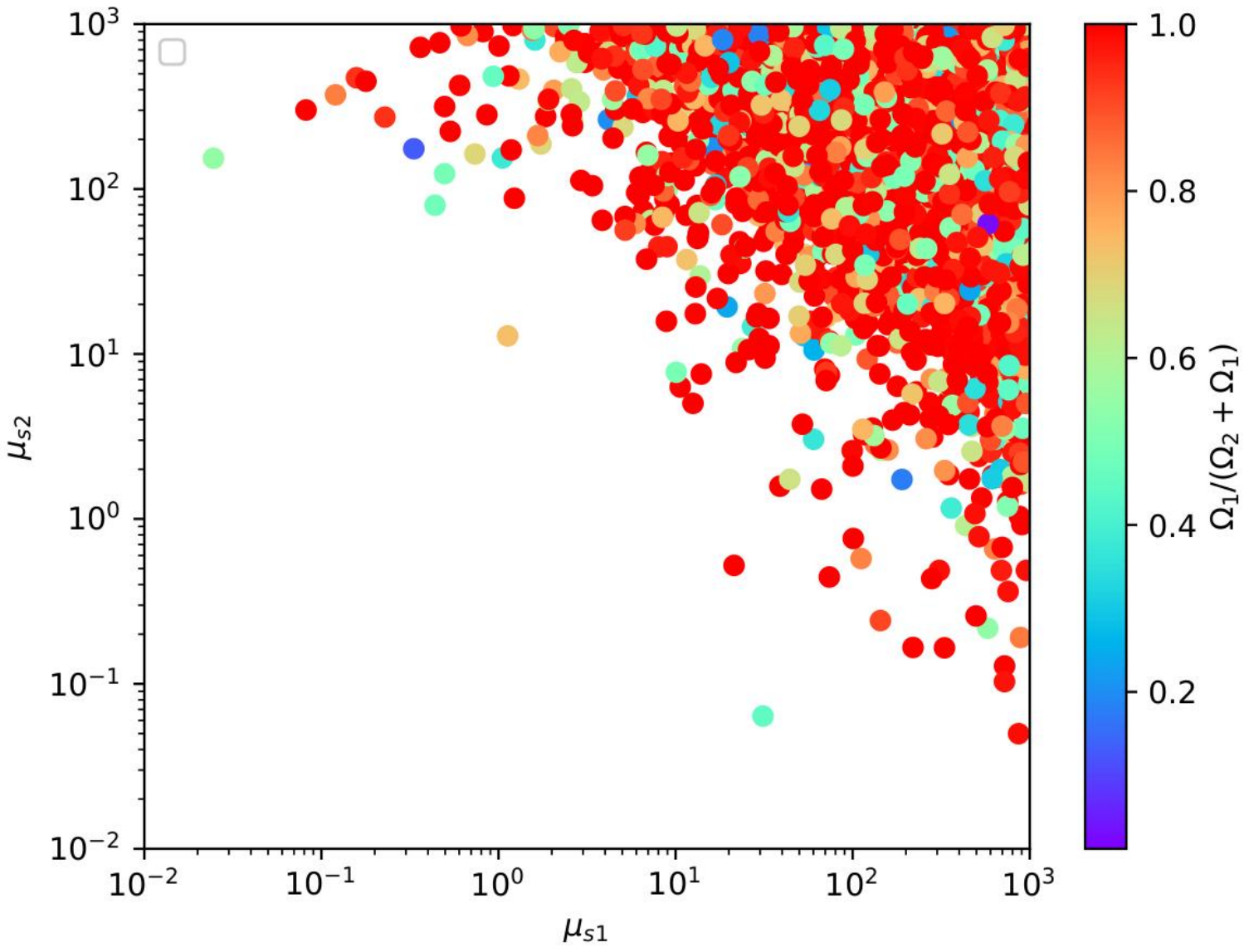}}
\subfigure[]{\includegraphics[height=5.5cm,width=6.5cm]{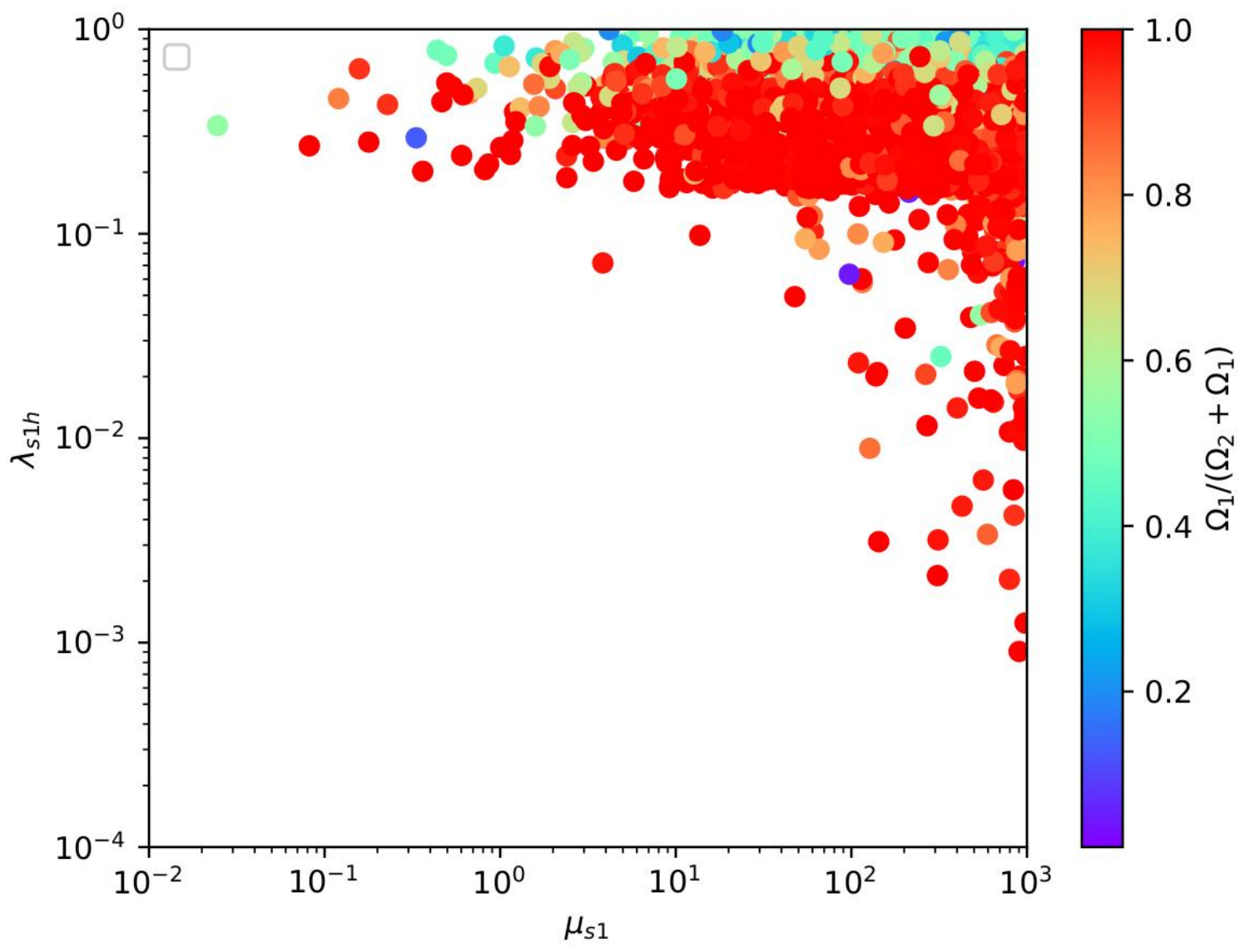}}
\caption{Parameter space satisfying relic density constraint, where points with different colors correspond to $\lambda_{412}$ in the picture (a) and (b), and the fraction $\Omega_1/(\Omega_1+\Omega_2)$ in the picture (c) and (d). The four pictures show the viable region of $\mu_{s1}-\lambda_{31}$, $\mu_{s2}-\lambda_{32}$, $\mu_{s1}-\mu_{s2}$ and $\mu_{s1}-\lambda_{s1h}$ respectively.}
\label{fig11}
\end{figure}
According to the latest experiment data measured by PLANCK \cite{Duda:2001ae}, the observed dark matter relic density is:
\begin{eqnarray}
 \Omega_{DM}h^2 = 0.1198 \pm 0.0012
\end{eqnarray}
This result will limit the parameter space of our model, and we make a random scan to consider a viable parameter space satisfying the dark matter relic density constraint between 0.11 and 0.13, which amounts to about a $10\%$ uncertainty. The parameters are varied in the following ranges:
\begin{eqnarray*}
 M_{\Delta} \subseteq [600,3000]\ \mathrm{GeV}, m_1 \subseteq [400,2500]\ \mathrm{GeV}, m_2 \subseteq [400,5000]\ \mathrm{GeV}, \mu_{si} \subseteq [0.01,1000]\ \mathrm{GeV}
 \end{eqnarray*}
 \begin{eqnarray}
 \lambda_{sih} \subseteq [10^{-4},1],\lambda_{sid} \subseteq [10^{-4},12.56], \lambda_{3i} \subseteq [10^{-4},1],\lambda_{412} \subseteq [0.01,1]. ~~(i=1,2)
\end{eqnarray}

In Fig.~\ref{fig8}(a), we show the viable parameter space of $M_{\Delta}-m_1$ satisfying relic density constraint, where points with different colors correspond to $m_2$ taking different colors. Similar to the case of two-component $Z_5$ dark matter in the SM, the whole mass region is allowed, and according to Fig.~\ref{fig8}(a), the larger $m_1$ always demands larger $m_2$ to satisfy dark matter relic density constraint. Unlike the two-component $Z_5$ dark matter in the SM, where the lighter component mainly determines dark matter relic density, we found both the heavy and light components can be dominant in the relic density in this work according to Fig.~\ref{fig8}(b). For most of the region, the fraction 
$\Omega_1/(\Omega_1+ \Omega_2)$ is close to 1, which means that the lighter component $\phi_1$ plays a dominant role in determining dark matter relic density. New annihilation processes related to the Type-II seesaw mechanism induce a wider parameter space in the model, and $\Omega_1/(\Omega_1+ \Omega_2)$ is in the range $(0,1)$. On one hand, the fraction is flexible in the case of $m_1 \approx m_2$, where the conversion processes related to $\phi_1$ and $\phi_2$ are suppressed, and both $\phi_1$ and $\phi_2$ can be dominant in dark matter relic density with proper parameters. On the other hand, the fraction can be more flexible for a larger $m_1$, where processes of dark matter annihilation into triplets can play an important role in determining dark matter relic density with the chosen parameters.

 From Fig.~\ref{fig9} to Fig.~\ref{fig11}, we give the viable parameter space of couplings satisfying relic density constraint. In Fig.~\ref{fig9}, we give the viable region of $\lambda_{s1h}-\lambda_{s1d}$, $\lambda_{s1h}-\lambda_{s2d}$ and $\lambda_{s2h}-\lambda_{s2d}$ respectively.  As we mentioned above, due to the existence of the triplet, the fraction $\Omega_1/(\Omega_1+\Omega_2)$ is flexible for the new annihilation processes. According to Fig.~\ref{fig9}(a),   almost all points lie on the upper-right region, which means for the small $\lambda_{s1h}$ and $\lambda_{s1d}$, the parameter space is excluded because of the too large relic density. Particularly, the allowed value for $\lambda_{s1h}$ is about $(10^{-3},12.56)$ and for the smaller $\lambda_{s1h}$, dark matter density will be over-abundant regardless of other parameter values. A larger $\lambda_{s1h}$ always demand a smaller $\lambda_{s1d}$ under relic density constraint. On the other hand, for the larger $\lambda_{s1h}$, the fraction is smaller which is consistent with the Freeze-out mechanism. We have similar conclusion for $\lambda_{s2h}$ and $\lambda_{s2d}$ according to Fig.~\ref{fig9}(b) and Fig.~\ref{fig9}(c), where $\lambda_{s2d}$ is constrained within $(10^{-3},12.56)$. The process $\phi_1 + \phi_1^{\dagger} \to \phi_2 + \phi_2^{\dagger}$ is determined by the combined contribution of Higgs-mediated diagrams as well as the quartic interaction, and the interference effects between the two diagrams may play a role so that result in either an increase or a decrease of the relic density. Therefore, we have a flexible parameter space for $\lambda_{s2h}-\lambda_{s2d}$ corresponding to different fraction $\Omega_1/(\Omega_1+\Omega_2)$ value.

 In Fig.~\ref{fig10} , we display the viable parameter space of $\lambda_{31}$, $\lambda_{32}$ and $m_1/m_2$. According to Fig.~\ref{fig10}(a) and Fig.~\ref{fig10}(b), most of the points lie in the upper-right region under dark matter relic density constraint, which means for the small $\lambda_{31}$ and $\lambda_{32}$, the corresponding interactions $\phi_1 + \phi_1 \to \phi_1 +\phi_2$  and $\phi_1 +\phi_2^{\dagger} \to \phi_2 + \phi_2$ is too small so that dark matter density is over-abundant. On the other hand, for the proper $\lambda_{31}$ and $\lambda_{32}$, one can always find corresponding parameters $m_1$ and $m_2$ satisfying dark relic density constraint and the fraction $\Omega_1/(\Omega_1+\Omega_2)$ can take value from $(0,1)$. In the Fig.~\ref{fig10}(c) and Fig.~\ref{fig10}(d), we give the relationship between $\lambda_{31}-m_1/m_2$ and $\lambda_{32}-m_1/m_2$, where points with different colors represent the fraction of $\phi_1$. For the smaller $m_1/m_2$, most of the points correspond to the fraction close to 1, and the total relic density is mainly composed of the lighter component $\phi_1$, which is consistent with the result of the two-component $Z_5$ dark matter in the SM. In the case of $m_1/m_2 \approx 1$, the fraction can take a value ranging from $(0,1)$, which means both the light and heavy components can determine dark matter relic density owing to the new annihilation channels. Note that the fraction $\Omega_1/(\Omega_1 +\Omega_2)$ can be small even for a small $m_1/m_2$ according to Fig.~\ref{fig10}(c), so the difference between $m_1$ and $m_2$ does not necessarily 
determine the component kind of dark matter relic density by tuning other proper parameters.

 According to Fig.~\ref{fig11}, we show the viable parameter space of $\mu_{s1}-\lambda_{31}$, $\mu_{s2}-\lambda_{32}$, $\mu_{s1}-\mu_{s2}$ and $\mu_{s1}-\lambda_{s1h}$ , where points with different colors correspond to $\lambda_{412}$ in Fig.~\ref{fig11}(a) as well as Fig.~\ref{fig11}  and represent $\Omega_1/(\Omega_1+\Omega_2)$ in Fig.~\ref{fig11}(c) and Fig.~\ref{fig11}(d) respectively. Process related to $\lambda_{31}$ ($\lambda_{32}$) is $\phi_1 +\phi_1 \to \phi_1 +\phi_2$ ($\phi_1 +\phi_2^{\dagger} \to \phi_2 + \phi_2$ ), but the trilinear coupling $\mu_{s1}$ ($\mu_{s2}$) can give rise to both semi-annihilation and conversion processes. Under relic density constraint, a larger $\mu_{s1}$ ($\mu_{s2}$) always demand a smaller $\lambda_{31}$($\lambda_{32}$), and with the increase of $\mu_{s1}$($\mu_{s2}$), the value of $\lambda_{31}$($\lambda_{32}$) is more flexible, which means contribution of $\phi_1 +\phi_1 \to \phi_1 +\phi_2$ ($\phi_1 +\phi_2^{\dagger} \to \phi_2 + \phi_2$) to relic density becomes less important for the large $\mu_{s1}$ ($\mu_{s2}$). On the other hand, dark matter density will be over-abundant for the small $\mu_{s1}(\mu_{s2})$ and $\lambda_{31}(\lambda_{32})$, therefore, most of the points satisfying relic density constraint lie in the upper-right region. The coupling $\lambda_{412}$
 determines the process $\phi_1+ \phi_1^{\dagger} \to \phi_2 +\phi_2^{\dagger}$ along with the couplings $\lambda_{s12}(\lambda_{s2h})$, and the interference effects between the Higgs-mediated diagrams and the quartic interaction $|\phi_1|^2|\phi_2|^2$ can result in either an increase or a decrease of the relic density. Hence the viable value of $\lambda_{412}$ is flexible according to Fig.~\ref{fig11}(a) and Fig.~\ref{fig11}(b). According to Fig.~\ref{fig11}(c), the fraction $\Omega_1/(\Omega_1+\Omega_2)$ can take value within $(0,1)$ for the allowed $\mu_{s1}-\mu_{s2}$, which means the dimensional couplings $\mu_{s1}$ and $\mu_{s2}$ will not influence the component of dark matter relic density significantly within the chosen parameter space. We show the relationship between $\mu_{s1}$ and $\lambda_{s1h}$ in the Fig.~\ref{fig11}(d) and the behavior of $\mu_{s1}-\lambda_{s1h}$ are similar with the above parameters since $\mu_{s1}$ and $\lambda_{s1h}$  determine different interaction cross section. They are negative correlations
 under relic density constraint.

 The Higgs portal interactions $\lambda_{sih}$ with $i=1, 2$ can contribute to the elastic scattering of the dark matter off nuclei in the model, which can put a stringent constraint on the parameter space. The expression of the spin-independent (SI) cross section can be given as follows\cite{Belanger:2020hyh}:
  \begin{eqnarray}
  \sigma_i= \frac{\lambda_{Si}^2}{4\pi}\frac{\mu_R^2m_p^2f_p^2}{m_H^4m_i^2}
    \label{ddeq}
  \end{eqnarray}
  where $i=1, 2$, $\mu_R$ is the reduced mass, $m_p$ is the proton mass, $m_H$ the SM Higgs mass and $f_p \approx 0.3$ is the quark content of the proton. Current experiments on the direct detection of dark matter can be found in \cite{XENON:2018voc,PandaX-4T:2021bab}, and the PandaX-4T experiments \cite{PandaX-4T:2021bab} put the most stringent constraint on the spin-independent dark matter.
  
 The annihilation of dark matter  in the late universe is so low that making little difference in the total dark matter abundance. Nevertheless, the effects
of such residual annihilations on the cosmic rays can be significant, which enable us detect dark matter indirectly. Indirect detection of dark matter is based on possible signatures of its annihilation such as energetic gamma rays in regions with significant DM density, e.g. the Galactic center, and the strictest bounds come from the Fermi-LAT\cite{McDaniel:2023bju} observation of 30 dSphs for 14.3 years.

 \section{Anti-proton spectrum and electron/positron excess}
\label{sec:4}
\begin{figure}[htbp]
\centering
\subfigure[]{\includegraphics[height=5.5cm,width=6.5cm]{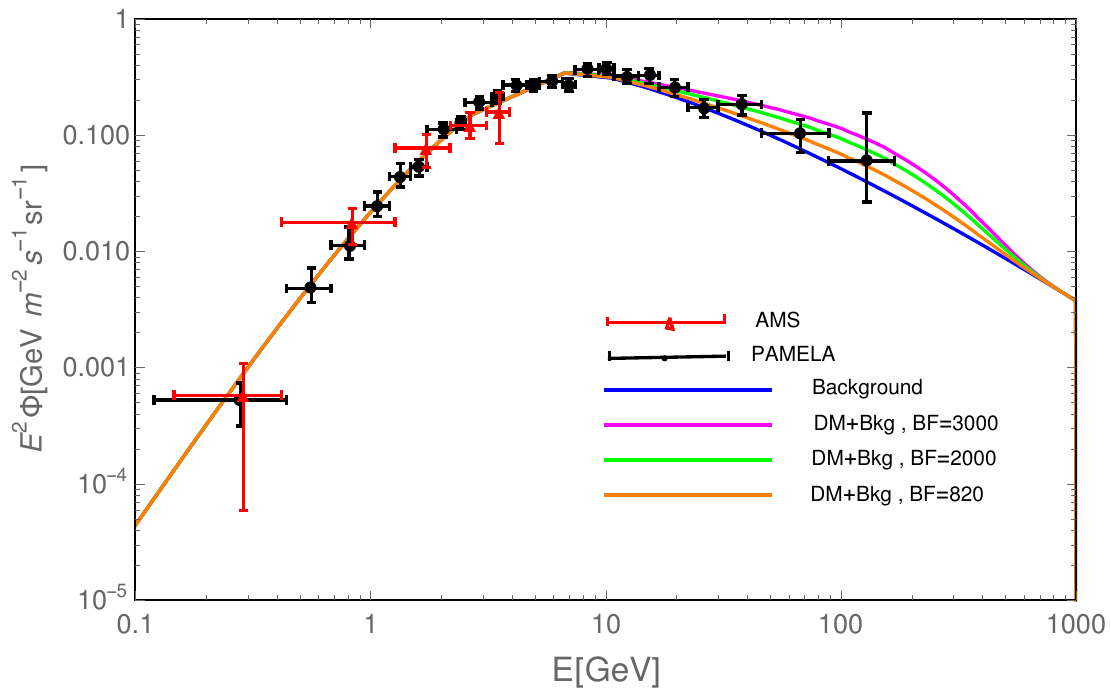}}
\subfigure[]{\includegraphics[height=5.5cm,width=6.5cm]{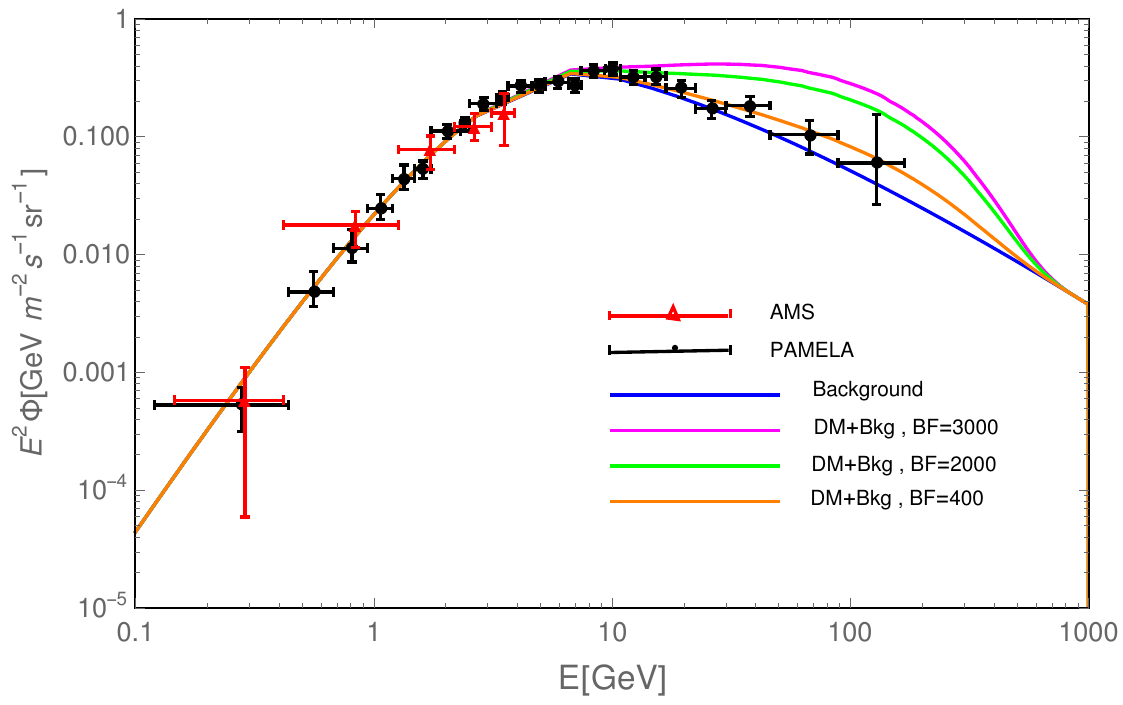}}
\subfigure[]{\includegraphics[height=5.5cm,width=6.5cm]{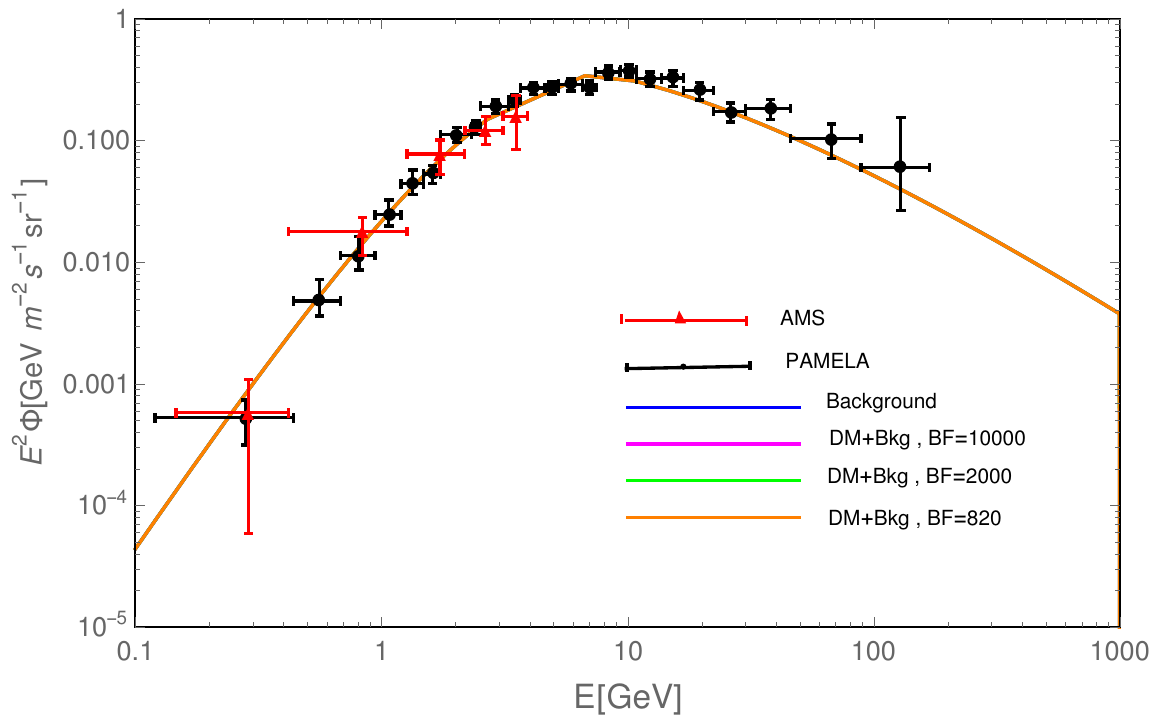}}
\subfigure[]{\includegraphics[height=5.5cm,width=6.5cm]{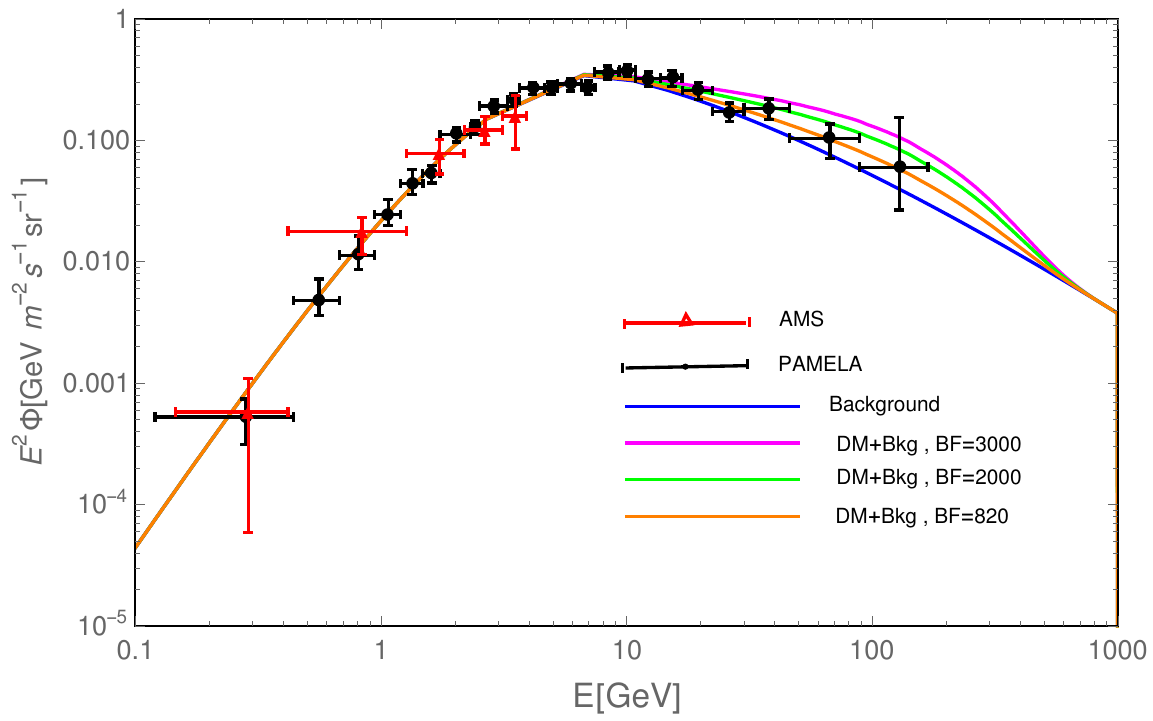}}
\subfigure[]{\includegraphics[height=5.5cm,width=8cm]{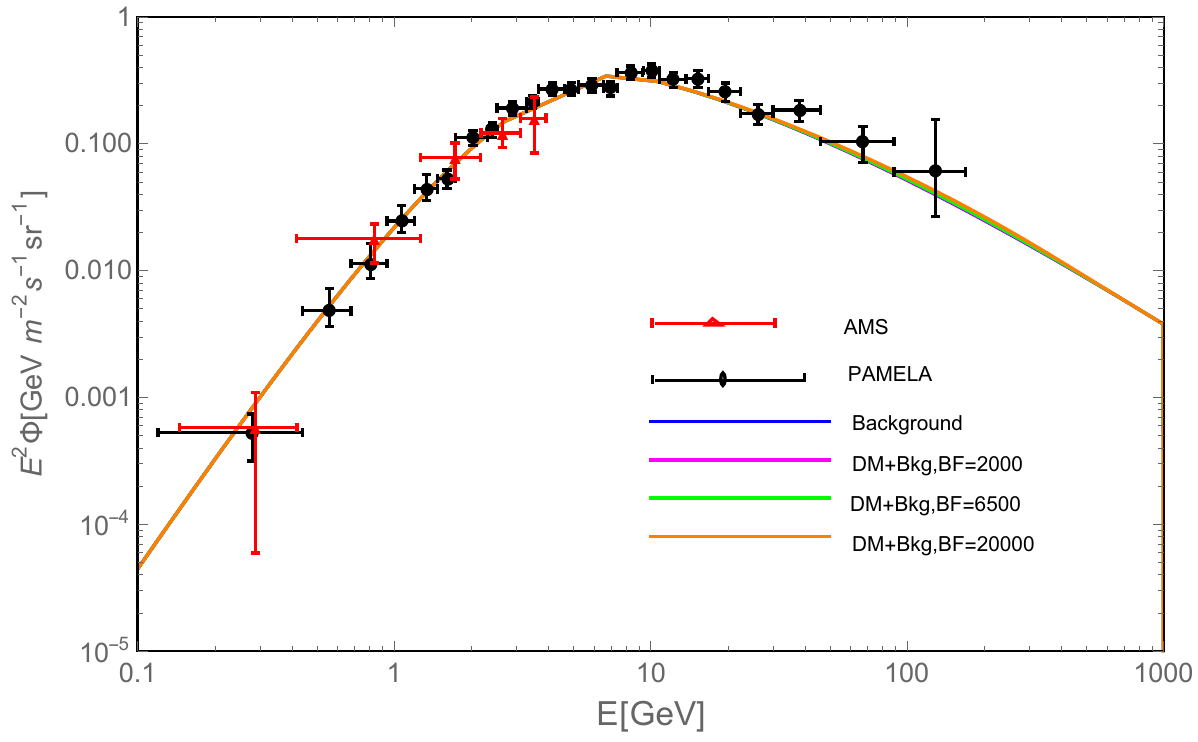}}
\caption{Background and background + DM (Bkg + DM) of the cosmic-ray antiproton flux with the five benchmark values respectively, where different colored lines correspond to the results of BF taking different values. The data points in both pictures label the PAMELA \cite{PAMELA:2010kea} and AMS \cite{AMS:2019nij} measurements.
}
\label{fig12}
\end{figure}
\begin{figure}[htbp]
\centering
\subfigure[]{\includegraphics[height=5.5cm,width=6.5cm]{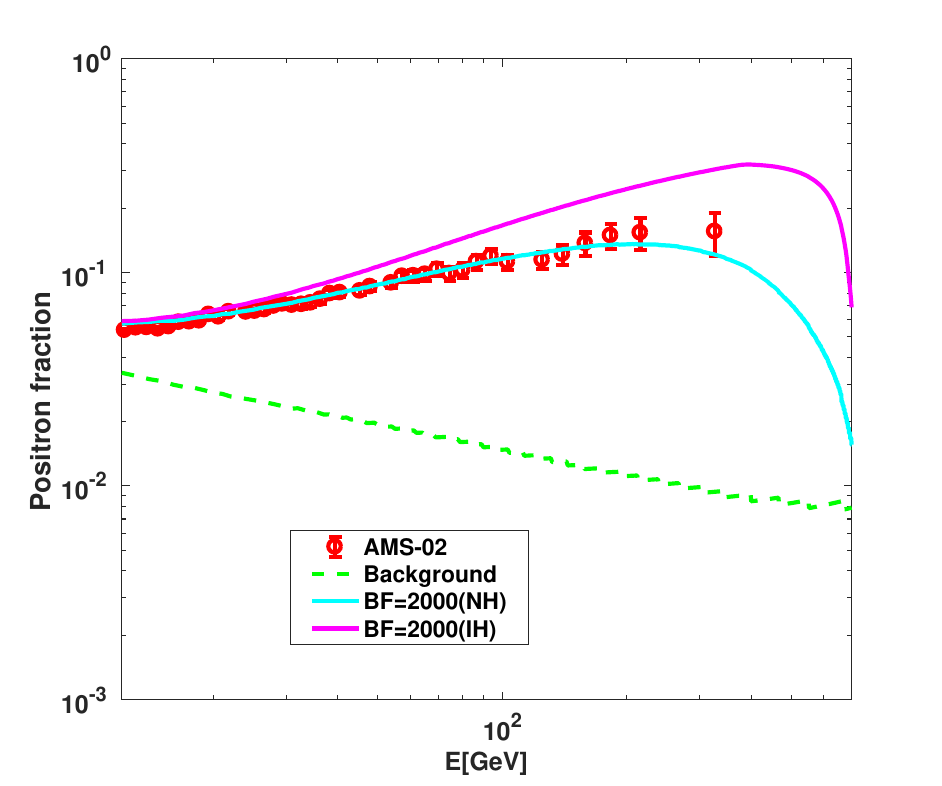}}
\subfigure[]{\includegraphics[height=5.5cm,width=6.5cm]{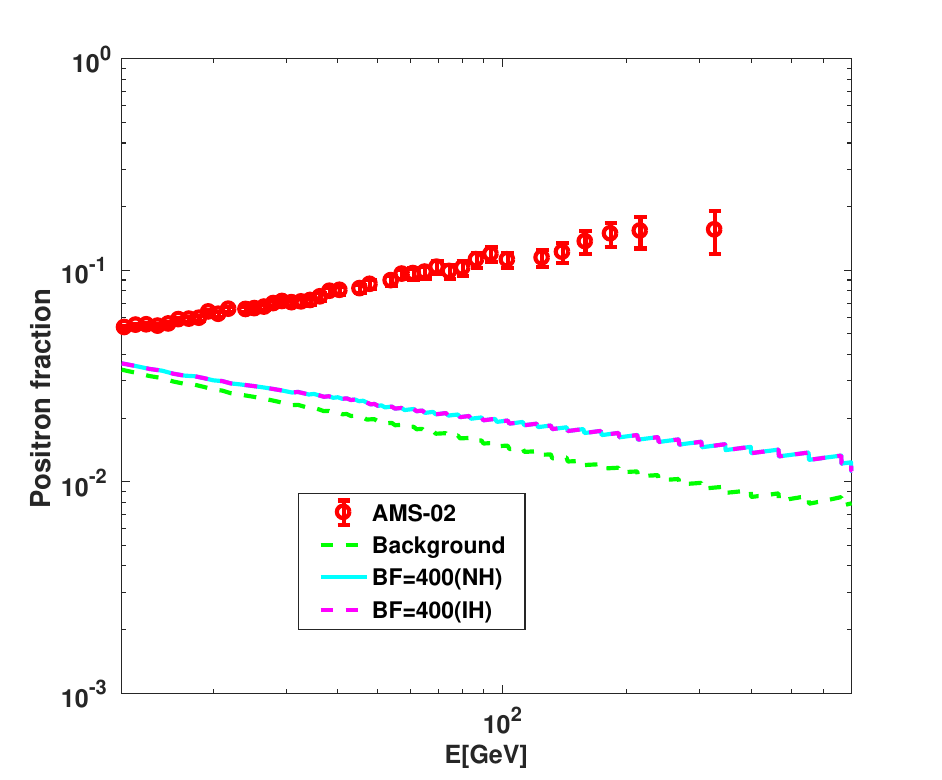}}
\subfigure[]{\includegraphics[height=5.5cm,width=6.5cm]{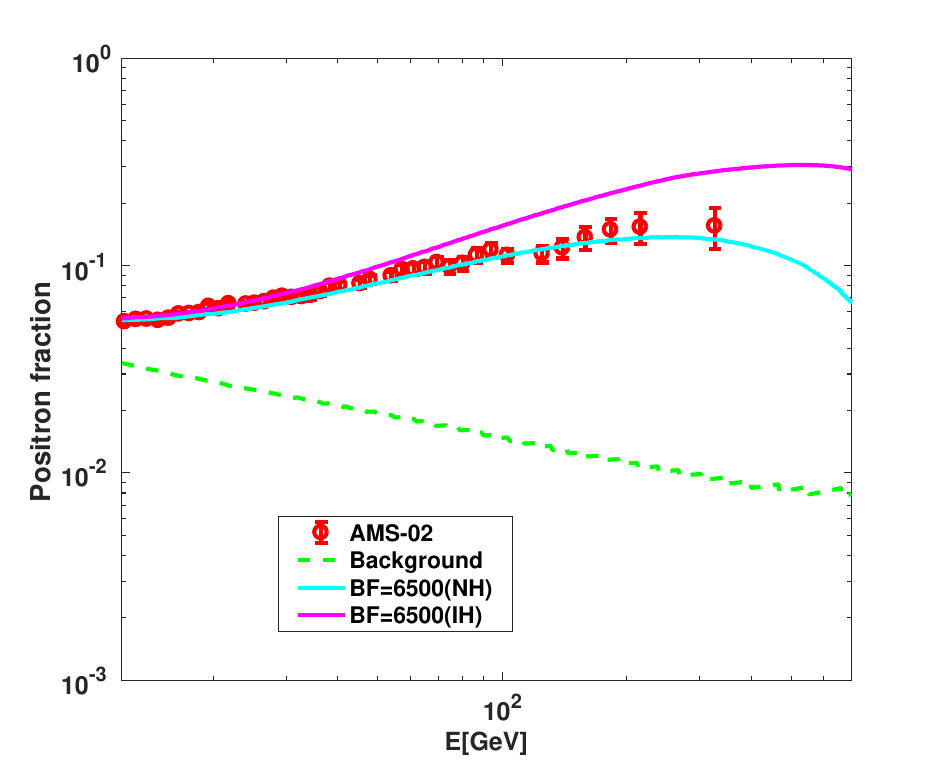}}
\subfigure[]{\includegraphics[height=5.5cm,width=6.5cm]{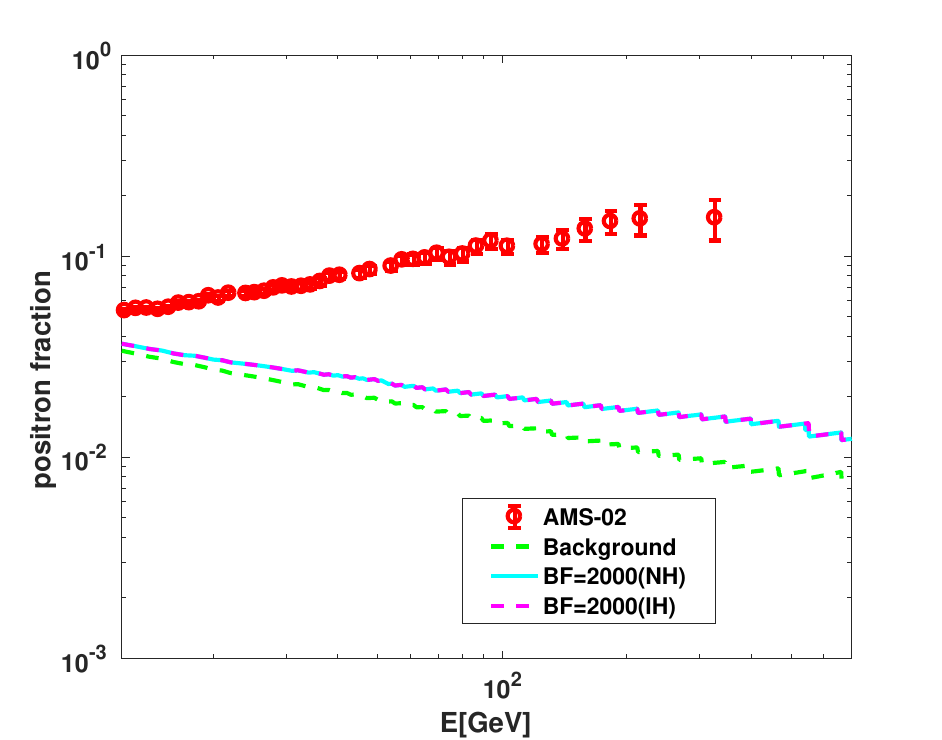}}
\subfigure[]{\includegraphics[height=5.5cm,width=8cm]{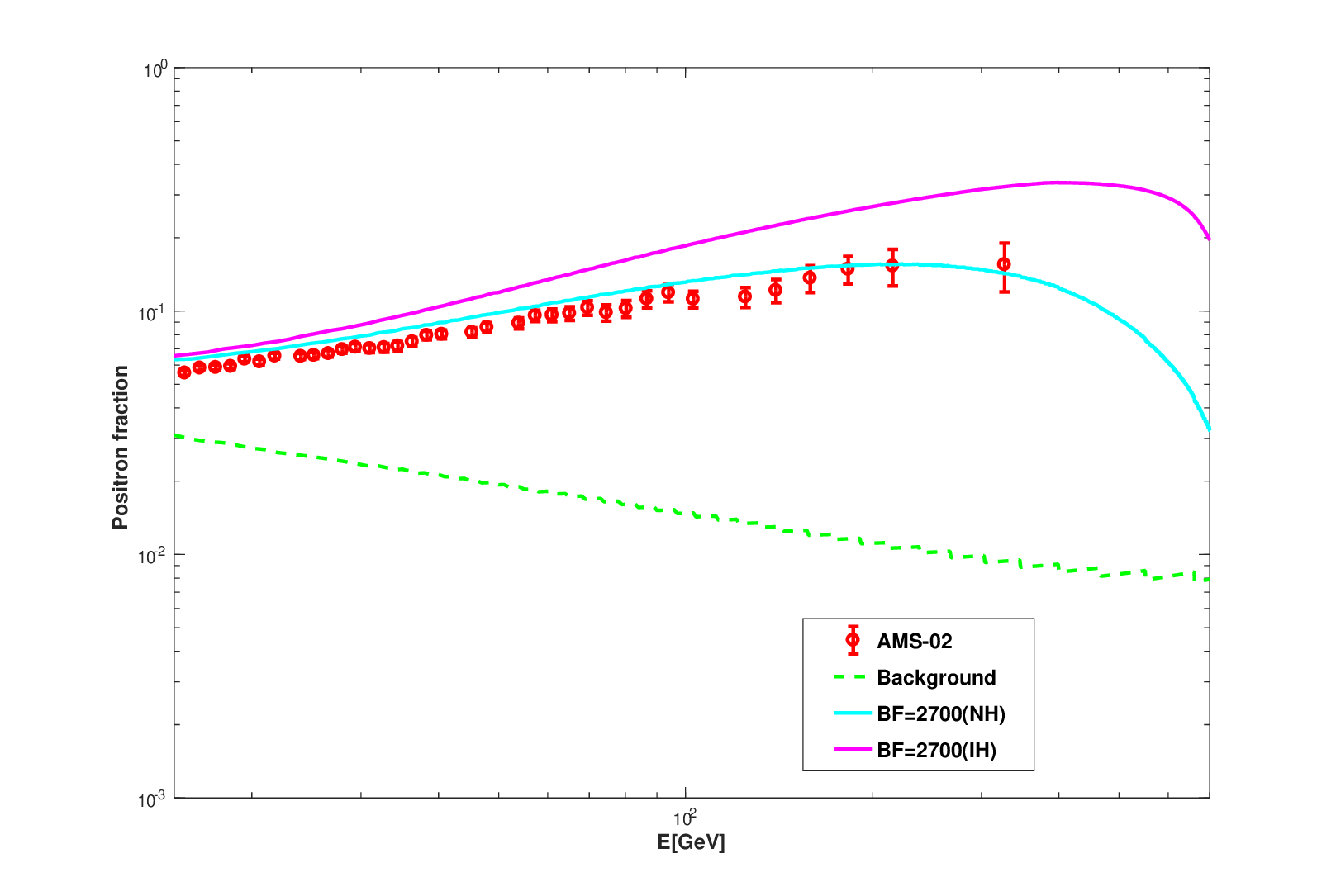}}
\caption{comparison of the positron fraction observed by
AMS-02 \cite{AMS:2014xys} in the IH scenario (the purple lines) and NH scenario (the blue lines) with the five benchmark values and fixed BF value. The data points stand for the AMS-02, and the green lines represent the Background.}
\label{fig13}
\end{figure}
 In the frame of Type-II seesaw, the triplets produced from DM (co)annihilation can decay to leptons ($\Delta \to l_il_j$) through Yukawa interactions, so that it is possible to account for the excess of electron-positron flux in cosmic rays exhibited in the AMS-02, Fermi-LAT, and DAMPE experiments. Note that the excess of positron-electron flux in cosmic rays may also be explained by astrophysical evidence, such as an isolated young pulsar \cite{Yuan:2017ysv}. Here in this paper, we focus on the DM interpretation, although it is likely that positron-electron 
fluxes are not due to DM but because of mundane astrophysics. Interpretation of the cosmic-ray excesses by the dark matter has been discussed a lot, and related works can be found in
refs.~\cite{Li:2018abw,Okada:2017pgr,Randall:2019zol,YaserAyazi:2019psw,Kachelriess:2019oqu,Feng:2019rgm,Cappiello:2018hsu,Yuan:2018rys,Chen:2017tva,Jin:2017qcv,Liu:2017rgs}.

We consider the following parametrization function to calculate the  antiproton flux and electron-positron flux \cite{Bringmann:2006im,Baltz:1998xv,Baltz:2001ir}:
\begin{eqnarray}
\label{fanti}
&&\log_{10}\Phi^{bkg}_{\bar{p}}=-1.64+0.07x-x^2-0.02x^3+0.028x^4, \\
\label{eflux1}
&&\Phi^{prim}_{e^-}(E)=\frac{0.16E^{-1.1}}{1+11E^{0.9}+3.2E^{2.15}}[GeV^{-1}cm^{-2}s^{-1}sr^{-1}], \\
&&\Phi^{sec}_{e^{-}}(E)=\frac{0.70E^{0.7}}{1+110E^{1.5}+600E^{2.9}+580E^{4.2}}[GeV^{-1}cm^{-2}s^{-1}sr^{-1}], \\
\label{eflux2}
&&\Phi^{sec}_{e^{+}}(E)=\frac{4.5E^{0.7}}{1+650E^{2.3}+1500E^{4.2}}[GeV^{-1}cm^{-2}s^{-1}sr^{-1}],
\end{eqnarray}
where $x=\log_{10}T/GeV$, the label $\Phi^{bkg}_{\bar{p}}$ corresponds to the cosmic-ray antiproton background. The label $\Phi^{prim(sec)}$ in Eq.(\ref{eflux1})-Eq.(\ref{eflux2}) means the primary (secondary) cosmic of electron or positron background.
The above formula is appropriate for the energy range 10-1000 GeV according to ref.~\cite{Baltz:1998xv}. The primary and secondary electron backgrounds are originated from supernova remnants and cosmic ray spallation in the interstellar medium, respectively.
The secondary positron background comes from primary protons colliding with other nuclei in the interstellar medium.
Therefore, the total antiproton, positron plus electron flux and positron flux are given as follows:
\begin{eqnarray}
&&\Phi_{\bar{p}}=\Phi^{bkg}_{\bar{p}}+BF\times \Phi^{DM}_{\bar{p}}. \\
&&\Phi_{e^+}+\Phi_{e^-}=k(\Phi^{(prim)}_{e^{-}}+\Phi^{(sec)}_{e^{-}}+\Phi^{(sec)}_{e^{+}})+BF \times (\Phi^{DM}_{e^{-}}+\Phi^{DM}_{e^{+}}), \\
&& \Phi_{e^+}=k\Phi^{(sec)}_{e^{+}}+BF\times \Phi^{DM}_{e^{+}}
\end{eqnarray}
where  $\Phi^{DM}$ is the corresponding flux from DM pair annihilation, and both $\phi_1$ and $\phi_2$ can contribute to the flux in our work. According to \cite{Dev:2013hka,Baltz:1998xv,Moskalenko:1997gh},  the normalization of the primary electron flux is undetermined
and parameterized by $k$.  The total electron + positron flux as well as the positron fraction are required to be consistent with the updated AMS-02 results according to \cite{Dev:2013hka}, and such requirement favors $k$ ranging among $[0.8,0.9]$ in the model. In this work, we consider $k = 0.9$ to fit the experimental data. The background fluxes can in principle be estimated, and 
 we use micrOMEGAs to calculate $\Phi^{DM}$, where the density distribution of DM in the galactic halo is taken from Navarro-Frenk-White (NFW) density, and the effects of galactic-charged particle propagation and solar modulation are considered.

  We focus on calculating the flux with E ranging among 0–1000 GeV considering the difference in the electron-positron spectrum between the DAMPE and Fermi-LAT measurements when $\mathrm{E} > 1$ TeV. On the other hand, we can have a poor fit when we include the low-energy data points below 15 GeV, but such discrepancies for $\mathrm{E}< 15$ GeV can be accounted for by uncertainties caused by solar modulation \cite{Asaoka:2001fv} and background flux uncertainties according to \cite{Dev:2013hka}. What's more, we will not use other datasets such as Fermi-LAT but AMS-02 for the positron fraction since AMS-02 data are much more precious.

 As we mentioned above, both the light and heavy components can be dominant in dark matter relic density, and the contribution of dark matter annihilation into triplet will be different under the two cases. In the following discussion, we choose the DM mass at the TeV scale and $M_{\Delta}= 1.05$ TeV. For other parameters, we consider the following benchmark values satisfying relic density as well as direct detection constraints to give a natural DM explanation of the cosmic-ray measurements:
 \begin{table}[t]
    \begin{tabular}{|c |c|}
     \hline
         &  Parameter values (i=1,2)\\
      \hline 
       Case I & \thead{$ m_1=1.1\ \mathrm{TeV}, m_2=1.3\ \mathrm{TeV}, \lambda_{412}=0.1, \mu_{si}=1\ \mathrm{TeV}$,\\ 
       $\lambda_{3i}=1, \lambda_{sih}=0.06, \lambda_{s1d}=0.68, \lambda_{s2d}=2$}\\
        \hline
        Case II  & \thead{$ m_1=1\ \mathrm{TeV}, m_2=1.37\ \mathrm{TeV},\lambda_{412}=0.1,\mu_{si}=1\ \mathrm{TeV}$, \\
        $\lambda_{31}=0.15,\lambda_{32}=0.18, \lambda_{s1h}=0.1,\lambda_{s2h}=0.01,\lambda_{s1d}=4.2,\lambda_{s2d}=0.1$}\\
        \hline
       Case III  & \thead{$ m_1=1.3\ \mathrm{TeV}, m_2=1.37\ \mathrm{TeV},\lambda_{412}=0.01,\mu_{s1}=1\ \mathrm{TeV}, \mu_{s2}=0.1\ \mathrm{GeV}$, \\
        $\lambda_{31}=1.45,\lambda_{32}=0.1, \lambda_{sih}=0.001, \lambda_{s1d}=5,\lambda_{s2d}=0.01$}\\
        \hline
       Case IV & \thead{$m_1=1.02\ \mathrm{TeV}, m_2=1.15\ \mathrm{TeV},\lambda_{412}=0.008,\mu_{si}=1\ \mathrm{TeV}$, \\
        $\lambda_{31}=0.6,\lambda_{32}=0.56, \lambda_{s1h}=0.12, \lambda_{s2h}=0.03, \lambda_{s1d}=10,\lambda_{s2d}=0.01$}\\
        \hline
        Case V & \thead{$m_1=1.1\ \mathrm{TeV}, m_2=1.21\ \mathrm{TeV},\lambda_{412}=0.01,\mu_{si}=1\ \mathrm{TeV}$, \\
        $\lambda_{31}=0.1,\lambda_{32}=0.3, \lambda_{s1h}=0.008, \lambda_{s2h}=0.008, \lambda_{s1d}=0.75,\lambda_{s2d}=1$}\\
        \hline
    \end{tabular}
    \caption{Benchmark parameters satisfying relic density constraint and direct detection constraint, where Case I (Case III), Case II (Case IV) correspond to the light (heavy) component dominant in dark matter relic density, and Case V is the case that the fraction of $\phi_1$ is approximately equal to $\phi_2$. 
    }
    \label{tab2}
\end{table}

 We show the results of the background and Bkg + DM for $\Phi_{\bar{p}} $ in Fig.~\ref{fig12} with the above five cases respectively, where the
PAMELA \cite{PAMELA:2010kea} results impose strong restrictions on the BF. In Fig.~\ref{fig12}(a) and Fig.~\ref{fig12}(b), we give the results of the light component dominant in dark matter relic density, while in Fig.~\ref{fig12}(c) and Fig.~\ref{fig12}(d), we give the heavy component cases. The allowed maximum values of  BF are about 2000,400 and 2000 respectively according to pictures (a),(b) and (d) of Fig.~\ref{fig12}. For the result of Case III (corresponding to Fig.~\ref{fig12}(c)), since the couplings $\lambda_{s1h} $ and $\lambda_{s2h}$ are so small that the contribution of  DM particles annihilating to W and Z boson pairs through the s-channel Higgs exchange can be negligible, therefore, the allowed BF value is more flexible compared with other cases, and we have similar results for Case V as in Fig.~\ref{fig12}(e), where the fraction of $\phi_1$ is approximately equal to $\phi_2$ is the total relic density.

In Fig.~\ref{fig13}, we give the corresponding results of the comparison of the positron
fraction observed by the AMS-02 with the five benchmark values, where the positron fraction is defined by $\Phi_{e^+}/(\Phi_{e^-}+\Phi_{e^+})$.  In the model, the final states of DM particle annihilation will depend on whether the neutrino mass is the inverted hierarchy (IH) or the normal hierarchy (NH). The annihilation production is mainly electron final states for the IH scenario, but tau final states for the NH scenario instead.  In Fig.~\ref{fig13}, the blue lines represent the results of the NH scenario and the purple lines correspond to the IH scenario.
When the triplets decay mainly into electron/positron final states, the positron flux arising from DM annihilation will rise sharply above the background as long as we go to higher energies and will eventually drop to the background level, while for the NH scenario, the triplets mostly decay into taus and subsequently decay to electrons, and one can obtain a much softer energy spectrum of the positron flux before eventually dropping to the background level as we can see in Fig.~\ref{fig13}(a), Fig.~\ref{fig13}(c) and Fig.~\ref{fig13}(e).  In Fig.~\ref{fig13}(b) and Fig.~\ref{fig13}(d) corresponding to Case II and Case IV, the results are above the background (without DM) but much lower than the AMS-02 data. The reason is that in those two cases, $m_1 <M_{\Delta}<m_2$, and the annihilation process of $\phi_1 + \phi_1^{\dagger} \to \Delta \Delta$ is kinetically suppressed, while the coupling $\lambda_{s2d}$ related to $\phi_2 + \phi_2^{\dagger} \to \Delta \Delta$ is so small that contribution of dark matter annihilation to triplet is also suppressed.  On the other hand, the NH scenario can give a better fit with the AMS-02 compared with the IH scenario as we can see in Fig.~\ref{fig13}(a), Fig.~\ref{fig13}(c) and Fig.~\ref{fig13}(e), which correspond to light component dominant, heavy component dominant and the fraction of $\phi_1$ approximately equal to $\phi_2$ in the model respectively.  As we mentioned above, the positron energy spectrum is harder in the IH scenario since it comes directly from the $\Delta$ decay, while the spectrum is softer in the NH case, because it comes from muon and tau decays.  The rise in the AMS-02 positron fraction becomes softer toward higher energies, therefore, it is easier to fit both low- and high-energy bins for the NH case compared to the IH case. In our model, both light components dominant in dark matter relic density and the heavy component case can give the proper results as we can see in Fig.~\ref{fig13}(a), Fig.~\ref{fig13}(c) and Fig.~\ref{fig13}(e) with $M_{\Delta}<m_1<m_2$, which is consistent with the results of dark matter in the Type-II seesaw mechanism. 
\section{summary and outlook}
\label{sec:so}
$Z_5$ two-component dark matter model has been discussed a lot, and one feature of such model is that dark matter relic density is mainly determined by the lighter component. In this work, we consider the $Z_5$ two-component dark matter model in the Type-II seesaw mechanism. We have new DM-scalar interactions so that both light and heavy components can be dominant in dark matter density with a wider parameter space satisfying relic density constraint. On the other hand, within the framework of the Type-II seesaw mechanism, the triplets and the W, Z boson pairs can be produced from DM (co)annihilation, and they can account for the interpretation of the excess of electron-positron flux as well as
antiproton spectrum with their subsequent decays.  Under such consideration, we discuss the cases of heavy and light components dominant within a viable parameter space satisfying relic density and direct detection constraints respectively, and in both cases, one can obtain the proper electron-positron flux excess with dark matter mass larger than triplets when neutrino mass is the normal hierarchy, which is consistent with the results of dark matter in the Type-II seesaw mechanism. Last but not least, as mentioned in \cite{Qi:2021rpa}, the semi-annihilation cross-section fraction is strongly constrained by the antiproton spectrum observed in the PAMELA and AMS experiments in the $Z_3$ scalar dark matter with Type-II seesaw mechanism. Similarly, 
there also exist other processes such as semi-annihilation processes and conversion processes in the model, and the possible influence of these different interactions on the electron-positron flux excess as well as antiproton spectrum can be interesting and we will discuss such effects in future work.
  \begin{acknowledgments}
\noindent
Hao Sun is supported by the National Natural Science Foundation of China (Grant No. 12075043, No. 12147205).
\end{acknowledgments}
\appendix
\section{Appendix}\label{appA}
\subsection*{FORMULAS}
The unitarity conditions come from the tree-level scalar-scalar scattering matrix which is
dominated by the quartic contact interaction. The s-wave scattering amplitudes should
lie under the perturbative unitarity limit, given the requirement the eigenvalues of the
S-matrix $\mathcal{M}$ must be less than the unitarity bound given by $|\mathrm{Re}\mathcal{M}| < \frac{1}{2}$.

The first submatrix ${\cal M}_1$ of the scattering whose initial and final states have charge zero: 
$E_1=($ $G^+\delta^-$, $\delta^+G^-$, $h\eta^0$, $\delta^0G^0$, $G^0\eta^0$, $h\delta^0$, $h\phi_1$, $\delta^0\phi_1$, $G^0\phi_1$, $\eta^0\phi_1$, $h\phi_1^*$, $\delta^0\phi_1^*$, $G^0\phi_1^*$, $\eta^0\phi_1^*$, $h\phi_2$, $\delta^0\phi_2$, $G^0\phi_2$, $\eta^0\phi_2$, $h\phi_2^*$, $\delta^0\phi_2^*$, $G^0\phi_2^*$, $\eta^0\phi_2^*$), and the submatrix $\mathcal{M}_1$ can be diveded into three block matrixs as follows:
\begin{equation}
{\scriptsize{\cal M}_1=
\left(
\begin{array}{cccccccc}
A&&&&\\
&&B&&\\
&&&&C\\
\end{array}
\right).}
\end{equation}
where A is a $6\times6$ matrix,B and C are $8\times 8$ matrixs  with:
\begin{equation}
{A=
\left(
\begin{array}{cccccc}
\lambda_1+\frac{\lambda_4}{2}&0&\frac{i\lambda_4}{2\sqrt{2}}&-\frac{i\lambda_4}{2\sqrt{2}}&\frac{\lambda_4}{2\sqrt{2}}&\frac{\lambda_4}{2\sqrt{2}}\\
0&\lambda_1+\frac{\lambda_4}{2}&-\frac{i\lambda_4}{2\sqrt{2}}&\frac{i\lambda_4}{2\sqrt{2}}&\frac{\lambda_4}{2\sqrt{2}}&\frac{i\lambda_4}{2\sqrt{2}}\\
\frac{i\lambda_4}{2\sqrt{2}}&-\frac{i\lambda_4}{2\sqrt{2}}&(\lambda_1+\lambda_4)&0&0&0\\
-\frac{i\lambda_4}{2\sqrt{2}}&\frac{i\lambda_4}{2\sqrt{2}}&0&\lambda_1+\lambda_4&0&0\\
\frac{\lambda_4}{2\sqrt{2}}&\frac{\lambda_4}{2\sqrt{2}}&0&0&\lambda_1+\lambda_4&0\\
\frac{\lambda_4}{2\sqrt{2}}&\frac{\lambda_4}{2\sqrt{2}}&0&0&0&\lambda_1+\lambda_4\\
\end{array}
\right).}
\end{equation}
\begin{equation}
{B=
\left(
\begin{array}{cccccccc}
\lambda_{s1h}&0&0&0&0&0&0&0\\
0&\lambda_{s1d}&0&0&0&0&0&0\\
0&0&\lambda_{s1h}&0&0&0&0&0\\
0&0&0&\lambda_{s1d}&0&0&0&0\\
0&0&0&0&\lambda_{s1h}&0&0&0\\
0&0&0&0&0&\lambda_{s1d}&0&0\\
0&0&0&0&0&0&\lambda_{s1h}&0\\
0&0&0&0&0&0&0&\lambda_{s1d}
\end{array}
\right).}
\end{equation}
\\
\begin{equation}
{C=
\left(
\begin{array}{cccccccc}
\lambda_{s2h}&0&0&0&0&0&0&0\\
0&\lambda_{s2d}&0&0&0&0&0&0\\
0&0&\lambda_{s2h}&0&0&0&0&0\\
0&0&0&\lambda_{s2d}&0&0&0&0\\
0&0&0&0&\lambda_{s2h}&0&0&0\\
0&0&0&0&0&\lambda_{s2d}&0&0\\
0&0&0&0&0&0&\lambda_{s2h}&0\\
0&0&0&0&0&0&0&\lambda_{s2d}
\end{array}
\right).}
\end{equation}
\\
\\
The second submatrix ${\cal M}_2$ of the scattering whose initial and final states have charge zero:
$E_2=($ $G^+G^-$, $\delta^+ \delta^-$, $\frac{G^0 G^0}{\sqrt{2}}$, $\frac{\eta^0 \eta^0}{\sqrt{2}}$,
$\frac{hh}{2}$, $\frac{\delta^0 \delta^0}{\sqrt{2}}$, $\delta^{++} \delta^{--}$, $\phi_1\phi_1^{*}$,$\phi_2\phi_2^{*}$),and these states have discrete $Z_5$ charge $X=0$. One can find:
\begin{equation}
{\scriptsize{\cal M}_2=
\left(
\begin{array}{ccccccccc}
4\lambda & \lambda_1+\frac{\lambda_4}{2}&\frac{4\lambda}{2\sqrt{2}}&\frac{\lambda_1}{\sqrt{2}}&\frac{4\lambda}{2\sqrt{2}}&\frac{\lambda_1}{\sqrt{2}}&\lambda_1+\lambda_4&\lambda_{s1h}&\lambda_{s2h}\\
\lambda_1+\frac{\lambda_4}{2}&4\lambda_2+2\lambda_3&\frac{2\lambda_1+\lambda_4}{2\sqrt{2}}&\sqrt{2}(\lambda_2+\lambda_3)&\frac{2\lambda_1+\lambda_4}{2\sqrt{2}}&\sqrt{2}(\lambda_2+\lambda_3)&2(\lambda_2+\lambda_3)&\lambda_{s1d}&\lambda_{s2d}\\\frac{4\lambda}{2\sqrt{2}} &\frac{2\lambda_1+\lambda_4}{2\sqrt{2}}&3\lambda&\frac{\lambda_1+\lambda_4}{2}&\lambda&\frac{\lambda_1+\lambda_4}{2}&\frac{\lambda_1}{\sqrt{2}}&\frac{\lambda_{s1h}}{\sqrt{2}}&\frac{\lambda_{s2h}}{\sqrt{2}}\\
\frac{\lambda_1}{\sqrt{2}}&\sqrt{2}(\lambda_2+\lambda_3)&\frac{\lambda_1+\lambda_4}{2}&3(\lambda_2+\lambda_3)&\frac{\lambda_1+\lambda_4}{2}&(\lambda_2+\lambda_3)&\sqrt{2}\lambda_2&\frac{\lambda_{s1d}}{\sqrt{2}}&\frac{\lambda_{s2d}}{\sqrt{2}}\\
\frac{4\lambda}{2\sqrt{2}}&\frac{2\lambda_1+\lambda_4}{2\sqrt{2}}&\lambda&\frac{\lambda_1+\lambda_4}{2}&3\lambda&\frac{\lambda_1+\lambda_4}{2}&\frac{\lambda_1}{\sqrt{2}}&\frac{\lambda_{s1h}}{\sqrt{2}}&\frac{\lambda_{s2h}}{\sqrt{2}}\\
\frac{\lambda_1}{\sqrt{2}}&\sqrt{2}(\lambda_2+\lambda_3)&\frac{\lambda_1+\lambda_4}{2}&(\lambda_2+\lambda_3)&\frac{\lambda_1+\lambda_4}{2}&3(\lambda_2+\lambda_3)&\sqrt{2}\lambda_2&\frac{\lambda_{s1d}}{\sqrt{2}}&\frac{\lambda_{s2d}}{\sqrt{2}}\\
(\lambda_1+\lambda_4)&2(\lambda_2+\lambda_3)&\frac{\lambda_1}{\sqrt{2}}&\sqrt{2}\lambda_2&\frac{\lambda_1}{\sqrt{2}}&\sqrt{2}\lambda_2&4(\lambda_2+\lambda_3)&\lambda_{s1d}&\lambda_{s2d}\\
\lambda_{s1h}&\lambda_{s1d}&\frac{\lambda_{s1h}}{\sqrt{2}}&\frac{\lambda_{s1d}}{\sqrt{2}}&\frac{\lambda_{s1h}}{\sqrt{2}}&\frac{\lambda_{s1d}}{\sqrt{2}}&\lambda_{s1d}&4\lambda_{41}&\lambda_{412}\\
\lambda_{s2h}&\lambda_{s2d}&\frac{\lambda_{s2h}}{\sqrt{2}}&\frac{\lambda_{s2d}}{\sqrt{2}}&\frac{\lambda_{s2h}}{\sqrt{2}}&\frac{\lambda_{s2d}}{\sqrt{2}}&\lambda_{s2d}&\lambda_{412}&4\lambda_{42}
\end{array}
\right).}
\end{equation}
\\
The third submatrix ${\cal M}_3$ of the scattering whose initial and final states have charge zero:
$E_3=($ $hG^0$, $\delta^0\eta^0$, $\frac{\phi_1\phi_1}{\sqrt{2}}$, $\frac{\phi_1^*\phi_1^*}{\sqrt{2}}$,$\frac{\phi_2\phi_2}{\sqrt{2}}$, $\frac{\phi_2^*\phi_2^*}{\sqrt{2}}$,$\phi_1\phi_2$,$\phi_1^{*}\phi_2^{*}$,$\phi_1\phi_2^{*}$,$\phi_1^{*}\phi_2$).
One can find:
\begin{equation}
{\scriptsize{\cal M}_3=\left(
\begin{array}{cccccccccc}
2\lambda&0&0&0&0&0&0&0&0&0\\
0&2(\lambda_2+\lambda_3)&0&0&0&0&0&0&0&0\\
0&0&\lambda_{41}&0&0&0&0&3\sqrt{2}\lambda_{31}&0&0\\
0&0&0&\lambda_{41}&0&0&3\sqrt{2}\lambda_{31}&0&0&0\\
0&0&0&0&\lambda_{42}&0&0&0&0&3\sqrt{2}\lambda_{32}\\
0&0&0&0&0&\lambda_{42}&0&0&3\sqrt{2}\lambda_{32}&0\\
0&0&0&3\sqrt{2}\lambda_{31}&0&0&\lambda_{412}&0&0&0\\
0&0&3\sqrt{2}\lambda_{31}&0&0&0&0&\lambda_{412}&0&0\\
0&0&0&0&0&3\sqrt{2}\lambda_{32}&0&0&\lambda_{412}&0\\
0&0&0&0&3\sqrt{2}\lambda_{32}&0&0&0&0&\lambda_{412}\\
\end{array}
\right).}
\end{equation}
\\

The forth submatrix ${\cal M}_4$ of the scattering whose initial and final states are charge one:
$E_4=($ $hG^+$, $\delta^0G^+$, $G^0G^+$, $\eta^0G^+$, $h\delta^+$, $\delta^0\delta^+$, $G^0\delta^+$,
$\eta^0\delta^+$, $\delta^{++}\delta^-$, $\delta^{++}G^-$, $\phi_1G^+$, $\phi_1\delta^+$, $\phi_1^*G^+$, $\phi_1^*\delta^+$,$\phi_2G^+$, $\phi_2\delta^+$, $\phi_2^*G^+$, $\phi_2^*\delta^+$), the submatrix ${\cal M}_4$ an be divided into three block matrixs with
\begin{equation}
{\scriptsize{\cal M}_4=
\left(
\begin{array}{cccccccc}
A&&&&\\
&&B&&\\
&&&&C\\
\end{array}
\right).}
\end{equation}
where A is a $10\times10$ matrix,B and C are $4\times 4$ matrixs  with:
\begin{equation}
{A=
\left(
\begin{array}{cccccccccc}
2\lambda&0&0&0&0&\frac{\lambda_4}{2\sqrt{2}}&0&-i\frac{\lambda_4}{2\sqrt{2}}&-\frac{\lambda_4}{2}&0\\
0&\lambda_1&0&0&\frac{\lambda_4}{2\sqrt{2}}&0&i\frac{\lambda_4}{2\sqrt{2}}&0&0&0\\
0&0&2\lambda&0&0&i\frac{\lambda_4}{2\sqrt{2}}&0&\frac{\lambda_4}{2\sqrt{2}}&-i\frac{\lambda_4}{2}&0\\
0&0&0&\lambda_1&-i\frac{\lambda_4}{2\sqrt{2}}&0&\frac{\lambda_4}{2\sqrt{2}}&0&0&0\\
0&\frac{\lambda_4}{2\sqrt{2}}&0&-\frac{\lambda_4}{2\sqrt{2}}&\frac{2\lambda_1+\lambda_4}{2}&0&0&0&0&-\frac{\lambda_4}{2}\\
\frac{\lambda_4}{2\sqrt{2}}&0&i\frac{\lambda_4}{2\sqrt{2}}&0&0&2(\lambda_2+\lambda_3)&0&0&-\sqrt{2}\lambda_3&0\\
0&i\frac{\lambda_4}{2\sqrt{2}}&0\frac{\lambda_4}{2\sqrt{2}}&0&0&0&\frac{2\lambda_1+\lambda_4}{2}
&0&0&-i\frac{\lambda_4}{2}\\
-i\frac{\lambda_4}{2\sqrt{2}}&0&\frac{\lambda_4}{2\sqrt{2}}&0&0&0&0&2(\lambda_2+\lambda_3)&-i\sqrt{2}\lambda_3&0\\
-\frac{\lambda_4}{2}&0&-i\frac{\lambda_4}{2}&0&0&-\sqrt{2}\lambda_3&0&-i\sqrt{2}\lambda_3&2(\lambda_2+\lambda_3)&0\\
0&0&0&0&-\frac{\lambda_4}{2}&0&-i\frac{\lambda_4}{2}&0&0&\lambda_1+\lambda_4\\
\end{array}
\right).}
\end{equation}
\begin{equation}
{B=
\left(
\begin{array}{cccc}
\lambda_{s1h}&0&0&0\\
0&\lambda_{s1d}&0&0\\
0&0&\lambda_{s1h}&0\\
0&0&0&\lambda_{s1d}\\
\end{array}
\right).}
\end{equation}
\\
\begin{equation}
{C=
\left(
\begin{array}{cccc}
\lambda_{s2h}&0&0&0\\
0&\lambda_{s2d}&0&0\\
0&0&\lambda_{s2h}&0\\
0&0&0&\lambda_{s2d}\\
\end{array}
\right).} 
\end{equation}
The fifth submatrix $M_5$ of the scattering whose initial and final states have charge two:
$E_5=($ $\frac{G^+G^+}{\sqrt{2}}$, $\frac{\delta^+\delta^+}{\sqrt{2}}$, $\delta^+G^+$, $\delta^{++}\delta^0$, $\delta^{++}\eta^0$,
$\delta^{++}G^0$, $\delta^{++}h$, $\delta^{++}\phi_1$, $\delta^{++}\phi_1^{*}$, $\delta^{++}\phi_2$, $\delta^{++}\phi_2^{*}$).
One can find:
\begin{equation}
{\scriptsize{\cal M}_5=
\left(
\begin{array}{ccccccccccc}
2\lambda&0&0&0&0&0&0&0&0&0&0\\
0&2\lambda_2+\lambda_3&0&-\lambda_3&-i\lambda_3&0&0&0&0&0&0\\
0&0&\frac{2\lambda_1+\lambda_4}{2}&0&0&-i\frac{\lambda_4}{2}&-\frac{\lambda_4}{2}&0&0&0&0\\
0&-\lambda_3&0&2\lambda_2&0&0&0&0&0&0&0\\
0&-i\lambda_3&0&0&2\lambda_2&0&0&0&0&0&0\\
0&0&-i\frac{\lambda_4}{2}&0&0&\lambda_1&0&0&0&0&0\\
0&0&-\frac{\lambda_4}{2}&0&0&0&\lambda_1&0&0&0&0\\
0&0&0&0&0&0&0&\lambda_{s1d}&0&0&0\\
0&0&0&0&0&0&0&0&\lambda_{s1d}&0&0\\
0&0&0&0&0&0&0&0&0&\lambda_{s2d}&0\\
0&0&0&0&0&0&0&0&0&0&\lambda_{s2d}\\
\end{array}
\right).}
\end{equation}
\\
The sixth submatrix ${\cal M}_6$ of the scattering whose initial and final states have charge three: $E_6=($ $\delta^{++}G^+$, $\delta^{++}\delta^+)$. 
All states have discrete $Z_5$ charge $X=0$. One can find:
\begin{equation}
{\scriptsize{\cal M}_6=
\left(
\begin{array}{cc}
\lambda_1+\lambda_4&0\\
0&2(\lambda_2+\lambda_3)
\end{array}
\right).}
\end{equation}
\\
The last submatrix ${\cal M}_7$ of the scattering whose initial and final states have charge four:
$E_7=\frac{\delta^{++}\delta^{++}}{\sqrt{2}}$, its discrete $Z_5$ charge $X=0$. 
The ${\cal M}_7$ is $2(\lambda_2+\lambda_3)$.

The eigenvalues $e^j_i$ of the submatrix ${\cal M}_i$ can be writen as
\begin{equation}
\begin{aligned}
&e^1_1=\lambda_1+\lambda_4,
e^2_1=\lambda_1,
e^3_1=\lambda_1+\frac{3}{2}\lambda_4,
e^4_1=\lambda_{s1h},\\
&e^5_1=\lambda_{s1d} ,
e^6_1=\lambda_{s2h},
e^7_1=\lambda_{s2d} ,
e^1_2=2\lambda,
e^2_2=2\lambda_2,
e^3_2=2(\lambda_2+\lambda_3), \\
&e^4_2=\lambda+\lambda_2+2\lambda_3+\sqrt{\lambda^2-2\lambda\lambda_2+\lambda^2_2-4\lambda\lambda_3+4\lambda_2\lambda_3+4\lambda^2_3+\lambda^2_4}    ,\\
&e^5_2=\lambda+\lambda_2+2\lambda_3-\sqrt{\lambda^2-2\lambda\lambda_2+\lambda^2_2-4\lambda\lambda_3+4\lambda_2\lambda_3+4\lambda^2_3+\lambda^2_4}    ,\\
&e^6_2=\frac{1}{4}Root[A],
e^1_3=2\lambda,
e^2_3=2(\lambda_2+\lambda_3),
e^3_3=\frac{1}{2} \left(-\sqrt{72 \lambda _{31}^2+\lambda _{41}^2+\lambda _{412}^2-2 \lambda _{41} \lambda _{412}}+\lambda _{41}+\lambda _{412}\right) ,\\
&e^4_3=\frac{1}{2} \left(\sqrt{72 \lambda _{31}^2+\lambda _{41}^2+\lambda _{412}^2-2 \lambda _{41} \lambda _{412}}+\lambda _{41}+\lambda _{412}\right),
e^5_3=\frac{1}{2} \left(-\sqrt{72 \lambda _{32}^2+\lambda _{42}^2+\lambda _{412}^2-2 \lambda _{42} \lambda _{412}}+\lambda _{42}+\lambda _{412}\right),\\
&e^6_3=\frac{1}{2} \left(\sqrt{72 \lambda _{32}^2+\lambda _{42}^2+\lambda _{412}^2-2 \lambda _{42} \lambda _{412}}+\lambda _{42}+\lambda _{412}\right),
e^1_4=\lambda_1+\lambda_4,
e^2_4=\lambda_1,\\
&e^3_4=\lambda_1+\frac{3\lambda_4}{2},
e^4_4=2\lambda,
e^5_4=2\lambda_2,
e^6_4=2(\lambda_2+\lambda_3),
e^7_4=\lambda_1-\frac{\lambda_4}{2}, \\
&e^8_4=\frac{1}{4}[4\lambda+4\lambda_2+8\lambda_3+\sqrt{(4\lambda-4\lambda_2-8\lambda_3)^2+16\lambda^2_4}],\\
&e^9_4=\frac{1}{4}[4\lambda+4\lambda_2+8\lambda_3-\sqrt{(4\lambda-4\lambda_2-8\lambda_3)^2+16\lambda^2_4}], \\
&e^{10}_4=\lambda_{s1h},
e^{11}_4=\lambda_{s1d},
e^{12}_4=\lambda_{s2h},e^{13}_4=\lambda_{s2d},
e^1_5=\lambda_1+\lambda_4 ,
e^2_5=\lambda_1,
e^3_5=2\lambda,
e^4_5=2\lambda_2,
e^5_5=2(\lambda_2+\lambda_3),\\
&e^6_5=\lambda_1-\frac{\lambda_4}{2},
e^7_5=2\lambda_2 -\lambda_3,
e^8_5=\lambda_{s1d},e^9_5=\lambda_{s2d},
e^1_6=\lambda_1+\lambda_4,
e^2_6=2(\lambda_2+\lambda_3),
e^1_7=2(\lambda_2+\lambda_3).
\end{aligned}
\end{equation}
Where we have ignored duplicate eigenvalues for ${\cal M}_i$. The symbol $Root[A]$ stand for the roots of cubic equation,
we will apply Sanmuelson's inequality \cite{1968How} to place restrictions on the region of roots.
\begin{align}
x^4&+(-24 \lambda -32 \lambda _2-24 \lambda _3-16 \lambda _{41}-16 \lambda _{42}) x^3+(-96 \lambda _1^2-96 \lambda _4 \lambda _1-24 \lambda _4^2-16 \lambda _{412}^2-48 \lambda _{s1d}^2-32 \lambda _{s1h}^2-48 \lambda _{s2d}^2\notag\\
&-32 \lambda _{s2h}^2+768 \lambda  \lambda _2+576 \lambda  \lambda _3+384 \lambda  \lambda _{41}+512 \lambda _2 \lambda _{41}+384 \lambda _3 \lambda _{41}+384 \lambda  \lambda _{42}+512 \lambda _2 \lambda _{42}+384 \lambda _3 \lambda _{42}+256 \lambda _{41} \lambda _{42}) x^2 \notag\\
&+(1536 \lambda _{41} \lambda _1^2+1536 \lambda _{42} \lambda _1^2+1536 \lambda _4 \lambda _{41} \lambda _1+1536 \lambda _4 \lambda _{42} \lambda _1-768 \lambda _{s1d} \lambda _{s1h} \lambda _1-768 \lambda _{s2d} \lambda _{s2h} \lambda _1+384 \lambda  \lambda _{412}^2+512 \lambda _2 \lambda _{412}^2 \notag\\
&+384 \lambda _3 \lambda _{412}^2+1152 \lambda  \lambda _{s1d}^2+768 \lambda _{42} \lambda _{s1d}^2+1024 \lambda _2 \lambda _{s1h}^2+768 \lambda _3 \lambda _{s1h}^2+512 \lambda _{42} \lambda _{s1h}^2+1152 \lambda  \lambda _{s2d}^2+768 \lambda _{41} \lambda _{s2d}^2+1024 \lambda _2 \lambda _{s2h}^2\notag\\
&+768 \lambda _3 \lambda _{s2h}^2+512 \lambda _{41} \lambda _{s2h}^2+384 \lambda _4^2 \lambda _{41}-12288 \lambda  \lambda _2 \lambda _{41}-9216 \lambda  \lambda _3 \lambda _{41}+384 \lambda _4^2 \lambda _{42}-12288 \lambda  \lambda _2 \lambda _{42}-9216 \lambda  \lambda _3 \lambda _{42}-6144 \lambda  \notag\\
& \lambda _{41}\lambda _{42}-8192 \lambda _2 \lambda _{41} \lambda _{42}-6144 \lambda _3 \lambda _{41} \lambda _{42}-384 \lambda _4 \lambda _{s1d} \lambda _{\text{s1h}}-384 \lambda _{412} \lambda _{s1d} \lambda _{s2d}-256 \lambda _{412} \lambda _{s1h} \lambda _{s2h}-384 \lambda _4 \lambda _{s2d} \lambda _{s2h}) x \notag\\
&+1536 \lambda _1^2 \lambda _{412}^2+384 \lambda _4^2 \lambda _{412}^2-12288 \lambda  \lambda _2 \lambda _{412}^2-9216 \lambda  \lambda _3 \lambda _{412}^2+1536 \lambda _1 \lambda _4 \lambda _{412}^2-18432 \lambda  \lambda _{42} \lambda _{s1d}^2-16384 \lambda _2 \lambda _{42} \lambda _{s1h}^2-\notag\\
&12288 \lambda _3 \lambda _{42} \lambda _{s1h}^2+1536 \lambda _{s1h}^2 \lambda _{s2d}^2-18432 \lambda  \lambda _{41} \lambda _{s2d}^2+1536 \lambda _{s1d}^2 \lambda _{s2h}^2-16384 \lambda _2 \lambda _{41} \lambda _{s2h}^2-12288 \lambda _3 \lambda _{41} \lambda _{s2h}^2-24576 \lambda _1^2 \lambda _{41} \lambda _{42}\notag\\
&-6144 \lambda _4^2 \lambda _{41} \lambda _{42}+196608 \lambda  \lambda _2 \lambda _{41} \lambda _{42}+147456 \lambda  \lambda _3 \lambda _{41} \lambda _{42}-24576 \lambda _1 \lambda _4 \lambda _{41} \lambda _{42}+12288 \lambda _1 \lambda _{42} \lambda _{s1d} \lambda _{s1h}+6144 \lambda _4 \lambda _{42} \lambda _{s1d} \lambda _{s1h}\notag\\
&+9216 \lambda  \lambda _{412} \lambda _{s1d} \lambda _{s2d}-3072 \lambda _1 \lambda _{412} \lambda _{s1h} \lambda _{s2d}-1536 \lambda _4 \lambda _{412} \lambda _{s1h} \lambda _{s2d}-3072 \lambda _1 \lambda _{412} \lambda _{s1d} \lambda _{s2h}-1536 \lambda _4 \lambda _{412} \lambda _{s1d} \lambda _{\text{s2h}}+\notag\\
&8192 \lambda _2 \lambda _{412} \lambda _{s1h} \lambda _{s2h}+6144 \lambda _3 \lambda _{412} \lambda _{s1h} \lambda _{s2h}+12288 \lambda _1 \lambda _{41} \lambda _{s2d} \lambda _{s2h}+6144 \lambda _4 \lambda _{41} \lambda _{s2d} \lambda _{s2h}-3072 \lambda _{s1d} \lambda _{s1h} \lambda _{s2d} \lambda _{s2h}=0
\end{align}
\bibliography {qxnew}
\end{document}